\documentclass[%
 aip,
 amsmath,amssymb,
 reprint,%
]{revtex4-1}

\newcommand{\nnm}{\nonumber}

\newcommand{\ra}{\rangle}
\newcommand{\la}{\langle}
\usepackage{color}
\usepackage[dvipdfmx]{graphicx}
\usepackage{epsfig}
\begin{document}

\title{Compact spin qubits using the common gate structure of fin field-effect transistors}%

\author{Tetsufumi Tanamoto}
\affiliation{Department of Information and Electronic Engineering, Teikyo University,
Toyosatodai, Utsunomiya  320-8511, Japan} 

\email{tanamoto@ics.teikyo-u.ac.jp}

\author{Keiji Ono}
\affiliation{Advanced device laboratory, RIKEN, Wako-shi, Saitama 351-0198, Japan}

\begin{abstract}
The sizes of commercial transistors are of nanometer order, and 
there have already been many proposals of spin qubits using conventional complementary metal–-oxide–-semiconductor (CMOS) transistors.
However, the previously proposed spin qubits  require many wires to control a small number of qubits.
This causes a significant 'jungle of wires' problem when the qubits are integrated into a chip.
Herein, to reduce the complicated wiring, 
we theoretically consider spin qubits embedded into fin field-effect transistor (FinFET) devices such that
the spin qubits share the common gate electrode of the FinFET.
The interactions between qubits occur via the Ruderman–-Kittel–-Kasuya–-Yosida (RKKY) interaction via the channel of the FinFET.
The possibility of a quantum annealing machine is discussed in addition to the 
quantum computers of the current proposals.
\end{abstract}

\maketitle
\section{Introduction}
Scalability and affinity with conventional computers are 
the most important features of semiconductor spin qubits~\cite{Loss0} when building 
a quantum circuit.
Recently, a number of significant developments have been achieved, 
greatly improving coherence time and fidelity~\cite{Carroll,Yoneda1,Yoneda2,Connors,Matias}.
The benefits of semiconductor qubits enable us to use the 
accumulated knowledge and technologies of 
the miniaturization of semiconductor devices, 
the gate lengths of which are already less than 20~nm in commercial use.
In this respect, the qubits based on current complementary metal–-oxide–-semiconductor (CMOS) field-effect transistor structures~\cite{Matias,Veldhorst,Maurand} 
have become more important in handling the trend of 
miniaturization of transistors.
However, 
it is questionable whether the qubit structures considered so far 
can be translated smoothly to mass production.
First, previous qubit structures require approximately ten electrodes to define, control, and read out a qubit.
This is because the direct qubit--qubit interaction requires a small distance 
between the two qubits, and the measurement structures are separated
from the qubit--qubit interaction parts. 
Although these setups have succeeded in a few qubit systems, 
if these qubits are to be integrated in a chip, the number of 
complicated wires will become a significant problem (referred to as the `jungle of wires' problem).
Moreover, when the qubit structures are far from the commercial base transistor 
architectures, a huge cost incurred in building the chips is unavoidable. 
The advanced nano-size transistors require several lithography masks 
via numerous complicated manufacturing processes~\cite{Liu}.
The high cost can only be made affordable 
if a large number of chips are expected to be sold in a large market, such as smartphones,
which would be far-future for the quantum computers because they currently only work at very low temperatures.
From this perspective, qubit structures should be as similar  
to those of conventional transistors as possible.

Herein, we theoretically investigate a compact spin-qubit system
embedded in common multi-gate FinFET transistors~\cite{BSIM,Fossum},
with all gates electrically tied together as the common gate.  
The quantum dots (QDs) as qubits are coupled with their nearest fin conducting channels.
The manipulations and measurements of the qubits are 
carried out by the common gate via the fin channels, 
in addition to the local magnetic fields across the qubits. 
The measurement is
described as a resonant behavior between the FinFET channel and QDs~\cite{Engel},
and it is shown that the 
energy difference between the nearest qubits is enhanced 
by the resonant structure.
Note that, because two channels couple with a QD and each channel other than the edge channel 
is shared with two QDs, the resonance is enhanced, resulting in the amplifying the 
detection of the qubit state.
By using the fin channels as the couplers between the qubits 
as well as the measurement current lines, 
the number of wires is greatly reduced and the 'jungle of wires' problem is solved.

\begin{figure*}
\centering
\begin{minipage}{17cm}
\includegraphics[width=12.0cm]{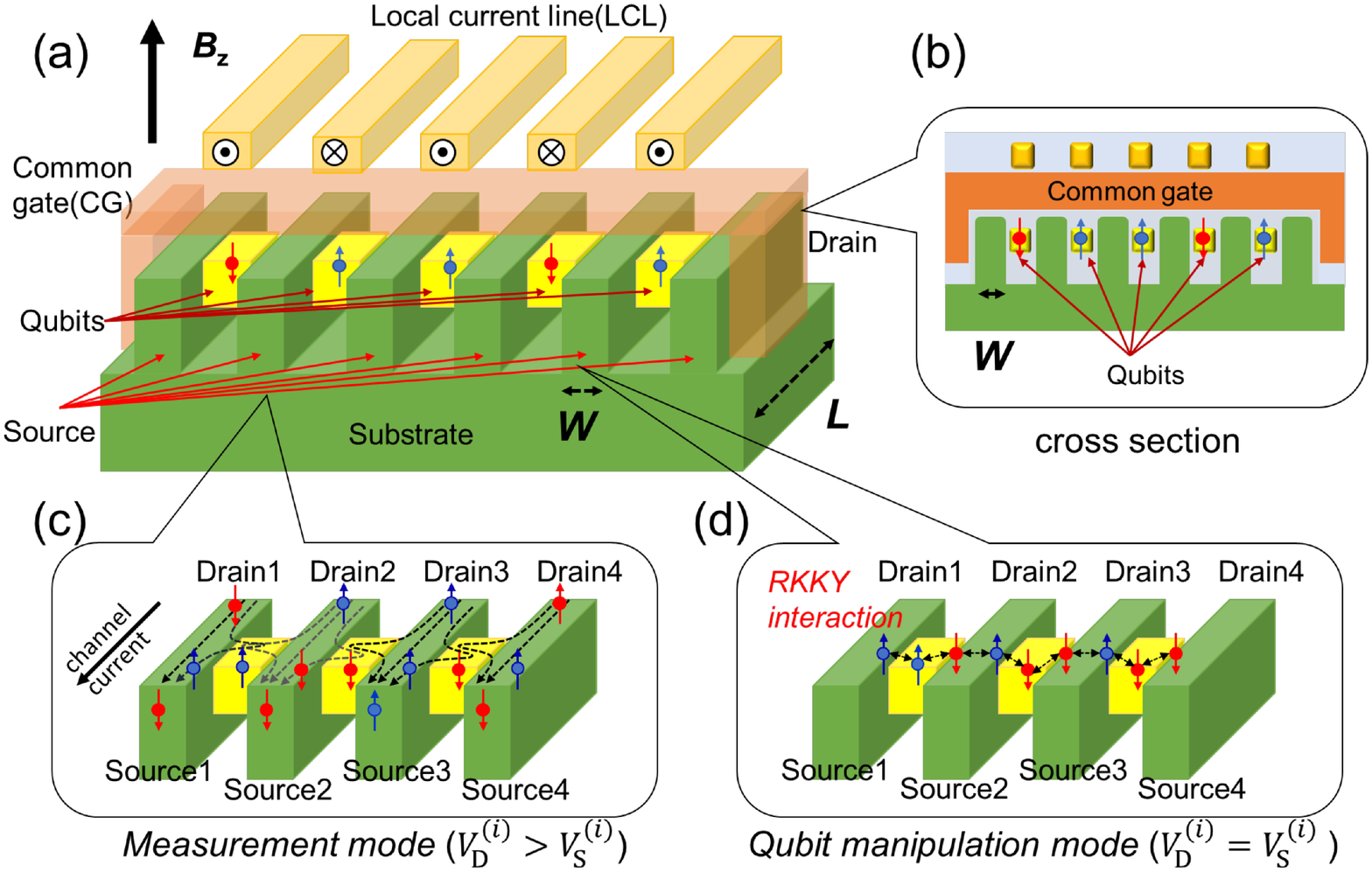}
\includegraphics[width=4.5cm]{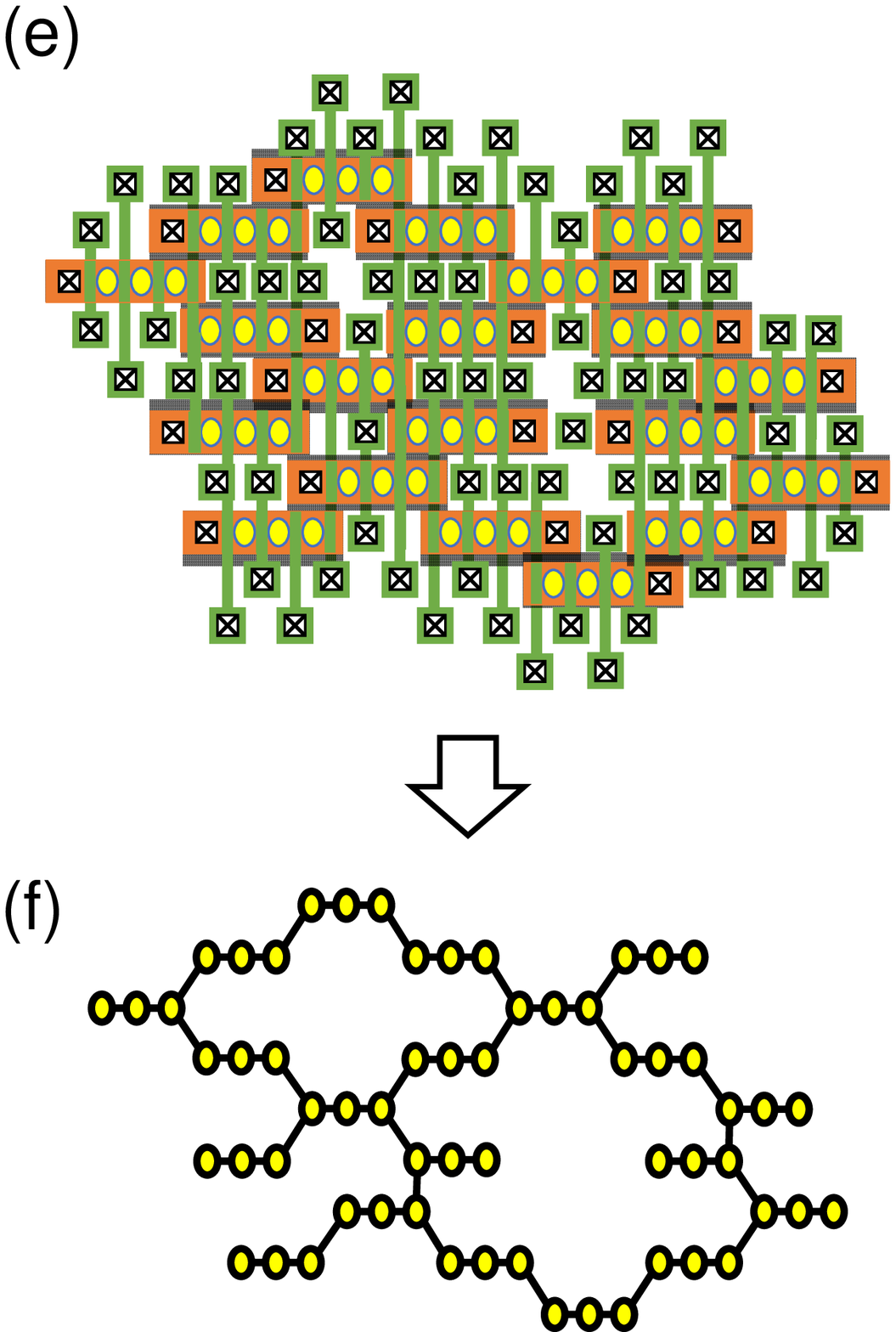}
 \end{minipage}
\caption{ Proposed common-gate spin qubits embedded in FinFET.
(a) Bird’s-eye view of the device. The electron spins in the quantum dots (QDs)
perform the role of qubits. The QDs are surrounded between neighboring conducting channels 
of a FinFET and constitute a spin chain across the device. 
The QDs share the gate electrode of the FinFET (the common-gate spin qubits).
The spin directions are controlled by the magnetic fields generated by the
`local current line (LCL)' over the common gate, and the external magnetic field.
The qubit states are detected by the FinFET conducting channel 
through the tunneling couplings between the QDs and the channels.
The $N$ qubits can be embedded into the $N+1$ fin channels.
The different  drains ($V_D^{(i)}$) and sources ($V_S^{(i)}$) are connected to the different channels ($i \leq N$), 
which also provide the difference from conventional FinFET devices.
(b) Cross-section of the proposed device perpendicular to the channel. 
(c)(d) The device has two operation modes. (c) shows the measurement mode 
in which FinFET currents flow from the source to the drain, interacting with the qubits via resonant tunneling.
The drain voltage is different from the source voltage. 
(d) shows the qubit manipulation mode in which qubits interact with each other via the channel charges 
(Ruderman–-Kittel–-Kasuya–-Yosida (RKKY) interaction). The currents do not flow ($V_D^{(i)}=V_S^{(i)}$).
Because the common gate structure is used, the complicated construction of gates is unnecessary.
The number $N$ depends on the target architecture. 
More than four fins should be combined to construct two-dimensional(2D) logic arrays  (e)(f). 
(e) Example of a 2D qubit array using the proposed spin qubit FinFET devices, and 
(f) the corresponding qubit network. 
The circles show the qubits, and the solid lines show the connections between the qubits.
}
\label{fig1}
\end{figure*}

\section{Results}
\subsection{Implementing qubits between the fin channels}
We start with the conventional FinFET structure.
FinFET types of transistors are widely used and can be extended to 
one-dimensional (1D) nanowires with gate lengths of less than 5~nm~\cite{imec2019}.
FinFET devices are developed to address the problems of 
orthodox planar CMOS transistors~\cite{BSIM,Fossum}.
Their ultra-thin bodies of less than 30 nm 
thickness enables them to solve the planar CMOS problem of leakage current 
between the source and drain.
The FinFET devices also solve the problem of random doping in the channel.
In addition, note that the thickness of the FinFET body ($<30$ nm) 
is less than that of the devices in previous spin-qubit experiments~\cite{Carroll,Yoneda1,Yoneda2,Connors,Matias}.

There are two choices of methods to embed the spin qubits into the FinFET device.
Lansbergen {\it et al.} located single-donor spin qubits in the channel of the FinFET device
~\cite{Lansbergen}. 
The other choice is to embed the spin qubits outside the channels.
The simplest structure is to array the spin qubits between the fin structure depicted in Fig.~\ref{fig1}.
The common gate structure is used in the same manner as the conventional FinFET.
A qubit is defined by an electron or hole in a QD (QDs can be replaced by trap sites~\cite{Ono}).
The source and drain electrodes are separated to 
detect the channel current independently.
The fabrication of this structure is within the scope of existing technologies~\cite{Karl}.
The excess charges are added to the QDs by biasing 
the two different channels surrounding the QDs.
The spin qubits are controlled by two orthogonal magnetic fields, $B_x$ and $B_z$.
The uniform magnetic field $B_z$ is applied to the sample, 
and the dynamic magnetic field $B_x$ is generated by the wires (LCL in Fig.~1) over the common gate.
The gate length $L$ and width $W$ are assumed to be less than 28 nm.
Both 2D and 1D channel electrons are assumed. 
The 2D electron gas is mainly formed on the surface of the channel structures
of the conventional FinFET. 
The 1D channel case corresponds to the nanowire FinFETs.

Because the qubits are spatially separated, 
the direct exchange interactions between the qubits cannot be used.
Instead, the interaction must be mediated by the channel electrons, 
and the RKKY interaction~\cite{Kittel,Kasuya,Yosida} 
using the channel electrons is the origin of the qubit--qubit interaction.
The RKKY interaction between two spin operators ${\bf S}_1$ and ${\bf S}_2$
is expressed by
\begin{equation}
H_{\rm RKKY}= J^{\rm RKKY}{\bf S}_1 \cdot {\bf S}_2,
\label{h_RKKY}
\end{equation}
where $J^{\rm RKKY}$ 
is the coupling constant between the qubits.
The tunneling of charges between the channel and the QDs forms the 
$s$--$d$ interaction between the spins in the QDs and those in the fin channel, 
and the RKKY interaction consists of the second-order 
perturbation of the $s$--$d$ interaction. 
Thus, the RKKY interaction is weaker than the direct-exchange interaction.  
As demonstrated later, the magnitudes of the RKKY interactions 
are estimated~\cite{Rikitake} as 0.01~meV and 0.2~$\mu$eV for 1D and 2D FinFET devices 
at the tunneling coupling energy $\Gamma$=0.15~meV for FinFETs of 28~nm and 14~nm gate length, respectively.
The corresponding coherence times are $10^{-9}\sim10^{-8}$~s.
The magnitude of the RKKY interaction depends on the Fermi level of the channel,
and the RKKY interaction is controlled by the 
applied gate voltage $V_G$ with source voltage $V_S^{(i)}$ and drain voltage $V_D^{(i)}$ ($i \in N$).

\begin{figure*}
\centering
\includegraphics[width=14cm]{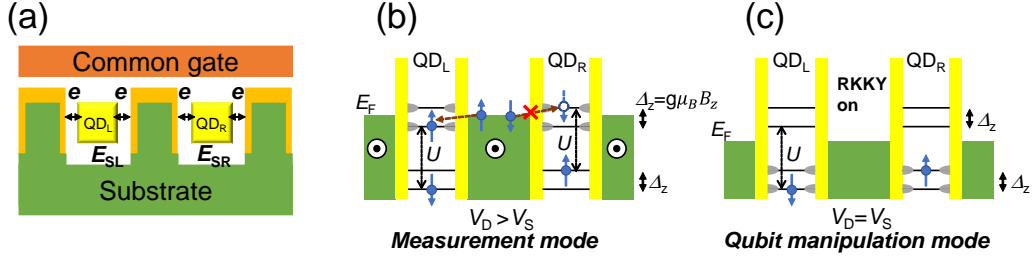}
\caption{Operation modes of the proposed spin qubits. 
(a) Two QDs and channel exchange electrons (a part of the multi-fin structure). 
(b, c) show the energy diagrams of the two operation modes.
The Coulomb interaction $U$ and Zeeman splitting $\Delta_z=g\mu_B B_z$ are assumed to be higher than 
the operation temperature. 
(b) Measurement mode. The drain voltage $V_D^{(i)}$ is larger than the source voltage $V_S^{(i)}$ $(i\in N)$.
The current through the channel is measured in a similar manner to that in a conventional transistor. 
(c) Qubit manipulation mode. 
The RKKY interaction is switched on when $V_D^{(i)}=V_S^{(i)}$, and 
its magnitude is controlled by the Fermi energy $E_F$ of the channel. 
$E_F$ is adjusted by $V_D^{(i)}$, $V_S^{(i)}$, and $V_G$.
}
\label{fig2}
\end{figure*}

\begin{figure}
\centering
\includegraphics[width=8.0cm]{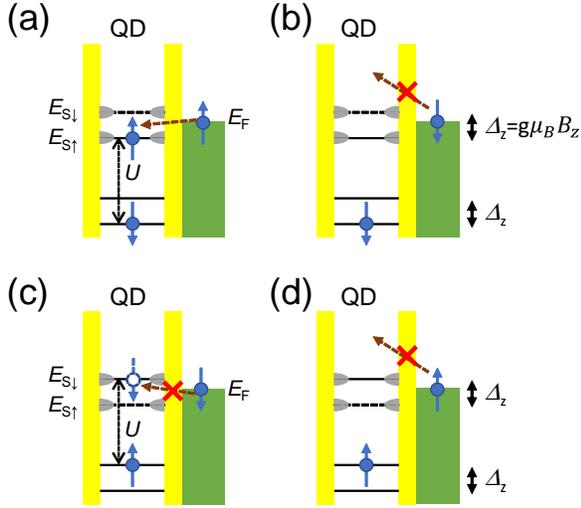}
\caption{Spin-filter effect in measurement mode. 
Spin-filter effects appear when the 
Fermi surface lies between the upper two energy levels.
(a) and (b) Spin down ($\downarrow$) state.
(c) and (d) Spin up ($\uparrow$) state.
The current flows only in state (a).
}
\label{figpattern}
\end{figure}

The two main operation modes (measurements, qubit manipulation)
are implemented by changing $V_G$, $V_S^{(i)}$ and $V_D^{(i)}$ (Figs.~1c and d).
The measurement mode and the qubit-manipulation mode are changed 
by the Fermi energy of the channel (Fig.~\ref{fig2}).
The qubit states are measured by the channel current of the FinFET devices.
The channel current reflects the spin up($\uparrow$) and down($\downarrow$) states of two QDs
when the Fermi energy lies between the upper two energy states (Fig.~\ref{figpattern}).
For example, when the upward magnetic field is applied to the device,
the current for the $\downarrow$-spin state is larger than that of the $\uparrow$-spin state  
(spin-filter effect).
The shot noise and thermal noise are analyzed and the signal-to-noise ratio is found to be larger than 100 if the applied magnetic field is sufficiently large.
In the quantum computing case, the idling mode is optional and discussed in the Appendix~\ref{Idling}.

The spin states are controlled by the local field $B_x$ and the global field $B_z$, 
in which the two qubit states ($\uparrow$-spin and $\downarrow$-spin) are 
distinguished by the Zeeman-energy splitting $g\mu_B B_z$ (hereafter we take $g=2$).
$B_x$ is generated by the currents of the LCLs over the gate electrodes (Fig.~1a).
Assuming that the distance $r$ between the QD and the LCL is $20$ nm,  
a magnetic field of $B_x=1$ mT is obtained
by the current 
$I=2\pi r B_x /\mu_{\rm Si} 
\approx 10 \mu$ A
for $\mu_{\rm Si}=10 \mu_0$ 
from Ampere’s law ($\mu_0= 1.26\times 10^{-6}$ kg~m$^{-2}$~s$^{-2}$A$^{-2}$).

Figures~1e and f show an example of the 2D qubit system and the corresponding qubit network.
Each FinFET can connect the qubits that belong to different FinFETs.
The magnitude of the RKKY interaction decreases with increasing distance between the qubits 
because of the Bessel function, as shown below.
Thus, the diagonal interactions between different FinFETs
are weaker than the interactions between neighboring qubits in the same FinFET.
Thus, the distances between different FinFETs should be minimized. 
Note that there are always strict design rules in the process technologies of each factory, 
and the distances cannot be shorter than fixed values. 
Here, we focus on a single FinFET device, and FinFET networks will be discussed in the near future.

In the case of general quantum computing, 
the global magnetic field is chosen as the quantized axis.
In this case, the spin direction is changed through the conventional rotating-wave approximation, and the frequency of the local field $\omega_i$ must 
satisfy $\hbar \omega \sim 2\mu_B B_z$~\cite{Ernst}. 
When we take a global field $B_z=1T$, we require $\omega \approx 28.0$ GHz, which can barely be transmitted to the local wire 
using present technologies.
If we deliver a 10 MHz local field, 
the corresponding Zeeman splitting is 6.58 peV, which cannot be measured.
At least a 10 GHz pulse is required for the Zeeman energy to be sufficiently separated.
High-frequency operation is required from the perspective of the coherence time mentioned above.
  
In the case of the quantum annealing machine (QAM)
~\cite{Nishimori,Finnila,Dwave,tana}, 
the quantized axis is generated by the LCL, and the 
uniformly applied external field is chosen as $B_x$.
The Hamiltonian is given by
$
H=\sum_{i<j} J_{ij}^{\rm RKKY} \vec{\sigma}_i \vec{\sigma}_j/4 
+ \sum_i [B_i^z \sigma_i^z + \Delta_i (t) \sigma_i^x]
$
($\sigma_i^\alpha$ ($\alpha=x,z$) are the Pauli matrices).  
The various data of the combinatorial problems are inputted 
into the RKKY interactions $J_{ij}^{\rm RKKY}$ and the local magnetic field $B_i^z$.
$J_{ij}^{\rm RKKY}$ is adjusted by the magnitudes of the Fermi energies 
of the fin channel, and  $B_i^z$ is adjusted by the LCLs.
In the present case, the Ising term is replaced by the Heisenberg
coupling term of equation~(\ref{h_RKKY}), 
and the tunneling term $\Delta (t) $ is produced by the global magnetic field.
The $\Delta (t)$ term is gradually switched off when the annealing process is complete. 
In this case, high-frequency operation 
of the magnetic field is not necessary but 
a large magnetic field should be produced by the LCL.
Note that there is a maximum current density to prevent electromigration for 
thin wires~\cite{ITRS}. 
Hu {\it et al.} investigated Cu wires with different cap materials for 7~nm and 14~nm transistors,
and demonstrated a reliable current density of 1.5 MA~cm$^{-2}$.
The wire with area 28~nm (width) $\times$ 56~nm (height) 
allows approximately 2.35~$\times$~10${}^{-4}$ A, which produces 
a $B_z^{(i)}$ of 23.5 mT ($\sim$2.722~$\mu$eV $\sim$31.5 mK),
assuming a distance of 20~nm between the LCL and the qubits.
This is small relative to the current possible operating temperature of 100~mK. 
If we can use 3 $\times 10^8$~A~cm$^{-2}$  NiSi nanowires~\cite{Wu},
the wire can generate a magnetic field of 470.4 mT ($\sim 54.5 \mu$eV $\sim$632.5~mK). 
Thus, use of the QAM is feasible 
if we can prepare reliable wires with current density greater than 100~MA~cm${}^{-2}$.
Thus, if the QDs are embedded between the FinFET devices, 
the 'jungle of wire' problem can be solved.
In the following, we describe the detailed analysis of our model. 


\subsection{Detailed analysis of the common-gate spin qubits}
Hereafter, we describe the theoretical detail of the common gate spin qubits.
In the FinFET device~\cite{BSIM,Fossum}, the carrier density of the fin channel can 
be changed within a range between 10${}^{15}$~cm${}^{-3}$ and 10${}^{19}$~cm${}^{-3}$,
by controlling the gate voltage within $0.3 \lesssim V_G \lesssim 1.2$ V.
Here, we consider the carrier density from the 10${}^{15}$~cm${}^{-3}$ to 10${}^{20}$~cm${}^{-3}$ region.
The corresponding Fermi energy 
$E_{Fd}$ 
of the 1D ($d=1$) and 2D ($d=2$) electron gas (hole gas) 
are estimated 
as  $0.188~{\rm meV} (75.2\mu {\rm eV})\lesssim E_{F1}\lesssim 0.405$~eV (0.162~eV), 
and $0.196~{\rm meV} (47.8\mu {\rm eV})\lesssim E_{F2}\lesssim 0.484$~eV (0.103~eV), respectively (see Appendix~\ref{parameters}).
The advantage of using the FinFET channel is that 
the adjustment of the gate bias $V_G$ enables us to control the Fermi energy of the channel, 
which leads to control of the measurement process and the qubit-- interaction.
We assume a Coulomb-blockade region of QDs 
where the charging energy is estimated as $U\approx 46.4$~meV for $L=W=10$~nm,
assuming a cubic QD of size $L_{\rm QD}=L/2$ (see Appendix~\ref{parameters}).
The discrete energy levels of the cubic QD $\epsilon_n$ are 
simply estimated by 
$\epsilon_n=\sum_{l=x,y,z}\pi^2\hbar^2 (n_l+1)^2/(2m^*L_{\rm Q}^2)$ 
($m^*$ is an effective mass, $n=\{n_x,n_y,n_z\}$ is an integer set where $n_l=0,1,..$), 
and we obtain $\epsilon_0 \approx 3.76$~meV 
for electrons, and 1.50~meV for holes. 
The corresponding energy of the first excited state $\epsilon_1$ is given as approximately 0.675~eV (0.270~eV), 
and we can consider single energy levels of the QDs 
(assuming that there is no offset to $\epsilon_0$ in the QDs). 
Hereafter, we consider the case of the electrons.
The two energy levels of the qubit state are defined by the $\uparrow$-spin state and the $\downarrow$-spin state 
under an external magnetic field $B_z$
in the resonant tunneling region~\cite{Ng}, such as  (Fig.~\ref{fig2}) 
\begin{equation}
T< J_{ij}^{\rm RKKY}<  2\mu_{\rm B} B_z  <E_F.
\label{condition}
\end{equation}

\begin{figure*}
\centering
\includegraphics[width=14.0cm]{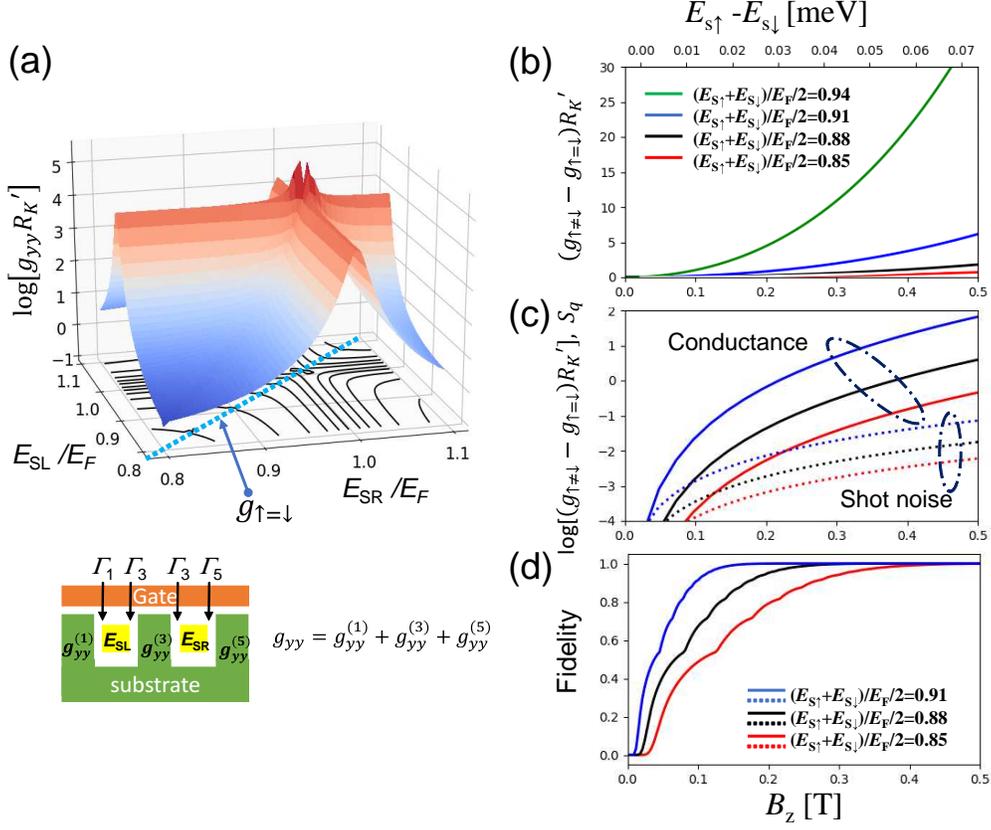}
\caption{Conductance characteristics of the proposed spin qubits. 
(a) Conductance (equation~(\ref{conductance})) as a function of the energy levels of the two QDs 
$E_{SL}$ and $E_{SR}$, where $E_{SL}$ and $E_{SR}$ are either $E_{S\uparrow}$ or $E_{S\downarrow}$.
$\Gamma_i/E_F=0.01$ ($i=1,3,5$) for simplicity.
$R_K'=(h/e^2/2)k_d=12.9k_d$k$\Omega$. $k_1=1$ (1D case) and $k_2= \pi n_{e2} W^2$ (2D case).
$n_{e2}$ is the number of carriers per nm$^{2}$.
(b) Difference in conductance as a function $E_{S\uparrow}-E_{S\downarrow}$, which is converted to the 
applied magnetic field $B_z$. 
$g_{\uparrow =\downarrow}$ is the conductance at $E_{SL}=E_{SR}$, and 
$g_{\uparrow \neq \downarrow}$ is that at  $E_{SL}\neq E_{SR}$ in Fig.~(a).
(c) Comparison between the conductance and shot noise.
The effect of the shot noise on conductance increases as the difference $E_{SR}-E_{SL}$ decreases.
(d) Measurement fidelity limited by the shot noise.
}
\label{figcond1}
\end{figure*}

\subsection{Measurement process}
The channel current reflects the QD states
when the Fermi energy of the channel is close to the energy levels of the QDs, as shown in Fig.~\ref{fig2}~b.
The positions of the upper energy levels are determined 
such that the upper and lower energy levels form the singlet states 
$|S\ra=[|\uparrow\downarrow\ra -|\downarrow\uparrow\ra]/\sqrt{2}$ 
(the triplet states 
are not considered 
because of their higher energy levels~\cite{Xuedong}).
The singlet energy state $E_{S\downarrow}$ for the $\downarrow$-state qubit    
is lower than that of the $\uparrow$-state qubit, 
given by $E_{S\downarrow}=E_{S\uparrow}-\Delta_z$.
Thus, as shown in Fig.~\ref{fig2}~b,
if we set the Fermi energy between $E_{S\uparrow}$ and $E_{S\downarrow}$,   
the $\uparrow$-spin electrons can tunnel between the QDs and the channel, 
but the $\downarrow$-spin electron tunneling is blocked (Fig.~\ref{figpattern}),
which is a spin filter effect similar to that in Ref.~\cite{Engel}.
The RKKY interaction is ineffective in this measurement mode,
because the conduction electrons that mediate 
the interaction between two QDs flow from the source to drain, 
and the RKKY interaction only works when both the $\uparrow$-spin and $\downarrow$-spin states 
are below the Fermi level.
Determination of the $\uparrow$-spin or $\downarrow$-spin state is performed by 
comparing the corresponding channel current with that of the reference channel current
in which both neighboring QDs have the same spin direction.  

Here, we analyze the conductance of the multi-channel FinFET device.
As a typical example, we consider the two QDs surrounded by three channels
as shown in Fig.~\ref{fig2}~a.
The basic setup is similar to that of the two-channel Kondo problem, 
except that we have to consider three current lines.
As Newns and Read~\cite{Read,Coleman} demonstrated, the standard approach to this 
problem is to apply the mean-field slave-boson approximation,
in which the number of electrons in the localized state is less than 1 and the spin-flop
process is included in the tunneling between the localized state 
and the conducting channel.
When $B_z$ is applied (Fig.~\ref{fig2}), 
the flip between $\uparrow$-spin and $\downarrow$-spin in the tunneling process is suppressed~\cite{Ng},
and we can investigate this setup in the range of the resonant-level model~\cite{Mahan}.
We assume that the scattering in the conducting channel is mainly caused by localized spins in the QDs.
All tunneling processes between the QDs and the fin channel are included.
In conventional FinFET circuit simulations, 
the drift-diffusion model is used as the core model to analyze the current characteristics~\cite{BSIM,Fossum}.
However, even in a conventional FinFET, 
more than 50\% of the current flows without scattering (ballistic transport)~\cite{Bufler}.
Thus, to examine the basic transport properties, we assume the scattering is caused only by the QDs. 

We derive the conductance using the Kubo formula~\cite{Kubo} based on the tunneling Hamiltonian
(see Appendix~\ref{LRT} section and supplement).
Figure~\ref{figcond1}a shows the conductance $g_{yy}$ of the summation of the 
three current lines  $g_{yy}^{(i)}$ ($i\in 1,3,5$) as a function of the energy levels of the two QDs,
where $E_{SL}$ and $E_{SR}$ are either 
$E_{S\downarrow}$ or $E_{S\uparrow}$.
We can observe a double-peak structure around the Fermi energy 
where $E_{SL}$ is close to $E_{SR}$ but $E_{SL} \neq E_{SR}$.
Because the double-peak structure can be observed even for a single fin channel ($g_{yy}^{(1)}=g_{yy}^{(5)}=0$, not shown),
we analyze the peak structure of the single channel $g_{yy}^{(3)}$.
The expression for $g_{yy}^{(3)}$ is 
\begin{equation}
g_{yy}^{(3)}
=\frac{2e^2}{h}k_d
\frac{4[\Delta^2 +\delta^2]^2}{
[(\Delta^2-2s_{33}\Delta -\delta^2)^2+4\Delta^2\Gamma_3^2]^2},
\label{conductance2}
\end{equation}
where $k_1=1$, $k_2= \pi n_{e2} W^2$,
$\Delta 
=(2E_{k_F}-E_{SL}-E_{SR}-s_{11}-s_{55})/2$, 
and $\delta
=E_{SL}-E_{SR}$. 
$n_{e2}$ is the number of the carriers per nm$^{2}$, and 
$s_{ij}\equiv \int |V_{\rm tun}(k_i)|^2/(E_{k_i}-E_{k_j})$ is the 
self-energy.
$\Gamma_i \approx 2\pi |V_{\rm tun}(k_i)|^2 \rho_F$ 
($\rho_F$ is the density of state at Fermi energy $E_F$, 
and $V_{\rm tun}$ is the overlap of wave functions 
between the channel and the QDs in the tunneling Hamiltonian).

The symmetric case $\delta=0$ gives the conventional 
resonant tunneling form $g =
4/[(\Delta-2s_{33})^2+4\Gamma_3^2]^2
$. 
In contrast, for the asymmetric case 
where $\delta\neq 0$ and $\Delta \ll \delta$, we have 
\begin{equation}
g_{yy} \rightarrow 4/\delta^4.
\end{equation}
Thus, the conductance increases 
as the asymmetry $\delta$ of the two QDs decreases 
for the region very close to the Fermi energy.
This is the origin of the sharp double peaks in Fig.~\ref{figcond1}a.
In general, realistic applications will require robustness to
variations in device parameters, 
and the double peaks might not be suitable for practical qubit detection
because they are sensitive to changes in $\{ E_{SL}, E_{SR} \}$. 
Instead, we consider 
the region where $\{ E_{SL}, E_{SR} \}$ are more distant from $E_F$ 
and the conductance changes gently.  
Figure~\ref{figcond1}~b shows the conductance changes
as a function of $E_{S\uparrow}-E_{S\downarrow}$,
the scale of which is converted to $B_z$. 
We can see that the conductance is approximately ten times larger than $R_K/2=2e^2/h$ 
($R_K\approx 25.8$~k$\Omega$ (von Klitzing constant)), 
which corresponds to the conductance of mS,
because $2e^2/h\approx 7.75 \times 10^{-5}$~S. 
Note that the transconductance of the FinFET is in the order of mS~\cite{BSIM,Fossum}. 
Thus, our results show that 
the FinFET devices can detect the energy 
difference $E_{S\uparrow}-E_{S\downarrow}$ of different qubits.
Moreover, because there are many fin channels, 
we can identify the spin direction of each qubit.
For example, in the case of the three fin channels, 
by setting $g_{yy}^{(1)} \neq 0$ and $g_{yy}^{(3)} =g_{yy}^{(5)}= 0$, 
it is possible to identify whether the left qubit is in the $\uparrow$-state or $\downarrow$-state. 
In general, for $N$ QDs and $N+1$ fin channels, 
the $i$th channel current is measured while $(i-1)$th and $(i+1)$th channel currents are switched off ($i<N$).
The $i-1$th and $i+1$th channel current can then be measured while the $i$th channel current is switched off.
By comparing the two cases, we can determine the spin directions of the $i-1$th and $i+1$th qubits.  
We can perform these processes in parallel to reduce the total measurement time.

We now consider the effect of noise.
There are unexpected trap sites in the FinFET devices.
Random telegraph noise (RTN) caused by capture and release of electrons 
at trap sites occurs in the order of $\mu$s~\cite{Grasser}. 
The RTN becomes a major problem when we consider a sequence of quantum algorithms,
because the voltage shift caused by the RTN is in the order of mV~\cite{Kamioka}.
Thus, we need to repeat the quantum operations to extract the 
desirable results.
Here, we focus on the shorter time region of two gate operations.
In this region, the shot noise and thermal noise are the main obstacles.
These types of FinFET device noise are in the order of $10^{-23}$ A$^2$~Hz$^{-1}$~\cite{BSIM}.
The shot noise is higher than the thermal noise (see Appendix~\ref{Noise}),
and its effect is described using our conductance formula.
The shot noise is given by 
$S_q= 2qI=2q g_{yy} V_D\sim 6.21\times 10^{-24}R_K g_{yy}$
for $V_D=0.5$~V. 
The conductance fluctuation originating from this shot noise is then given by
$
\Delta g_{yy} = \sqrt{ S_q \Delta f}/V_D =\sqrt{ 2q g_{yy} \Delta f/V_D}
$.
The condition $g_{yy} >\Delta g$ leads to $g_{yy} >2q \Delta f/V_D$.
Figure \ref{figcond1}c shows the comparison of $\Delta g_{yy}$ with the conductance difference.
As can be seen from this figure, the effect of the 
shot noise is small at qubit energy levels close to the Fermi energy.
Figure \ref{figcond1}d estimates the fidelity caused by the shot noise (see Appendix~\ref{Noise}).
The required $B_z$ decreases as the energy levels approach $E_F$.

\subsection{RKKY interaction and coherence time}
The physics regarding the coupling between localized-state and conduction electrons 
has a long history as the Kondo effect~\cite{Kondo}, other than the RKKY interaction~\cite{Sasaki,Marcus}.
The Kondo effect is observed below the Kondo temperature $T^K$.
In the Kondo regime $T<T^K$, the localized electrons in the QDs and 
channel electrons are coherently coupled and 
the initial qubit state is lost. 
Therefore, the Kondo effect is undesirable in our system.
For the RKKY interaction to be used effectively, 
the energy scale of the RKKY interaction should be larger than $T^K$~\cite{Doniach,Coleman1},
and the target parameter region is given by $J^{\rm RKKY}>T^K$.
The present setup is similar to the two-channel Kondo case.
Experimentally, it appears to be more difficult to observe the two-channel Kondo effect
than the single-channel Kondo effect~\cite{Potok,Zhu}. 
Here, we numerically compare the Kondo effect with the RKKY interaction.

\begin{table}
\caption{
The RKKY interaction $J^{\rm RKKY}$ and  the decoherence rate $\gamma^{\rm RKKY}$ are expressed for 1D ($d=1$) 
and 2D ($d=2$).
$F_d'(k_F W)$ consists of the Bessel functions $J_n(x)$ and $N_n(x)$.
$z_d\equiv \Gamma U/[(U-E_m)E_m]$ (see text). 
We assume that the area of a conducting electron $S$ includes 
two lateral planes and a top plane, and $S=L(W+2HFIN)$ with $HFIN=30$~nm.
$n_{e1}$ is the number of carriers per nm, and 
$n_{e2}$ is the number of carriers per nm$^{2}$.
}
\label{table:RKKY}
\begin{center}
\begin{tabular}{l|l|l|l|l}\hline
d  &$J^{\rm RKKY}_d$& $\gamma^{\rm RKKY}_d$ 
 & $k_F$  & $F_d'(x)$ \\ \hline
1 
& $\frac{z_1^2 E_F}{\pi} F_1'(k_F W) $ 
& $\frac{2 z_1^2 k_B T}{\pi}$  
& $\pi n_{e1}$ 
& ${\rm si}(2x)$
 \\ \hline
2  
& $\frac{z_2^2 E_F}{4\pi^3} F_2'(k_F W)$
& $\frac{z_2^2 k_B T}{8\pi^2}$   
& $\sqrt{2\pi n_{e2}}$ 
& $J_0(x)N_0(x)+J_1(x)N_1(x)$
 \\ \hline
\end{tabular}
\end{center}
\end{table}

\begin{figure*}
\centering
\begin{minipage}{17.5cm}
\includegraphics[width=9.0cm]{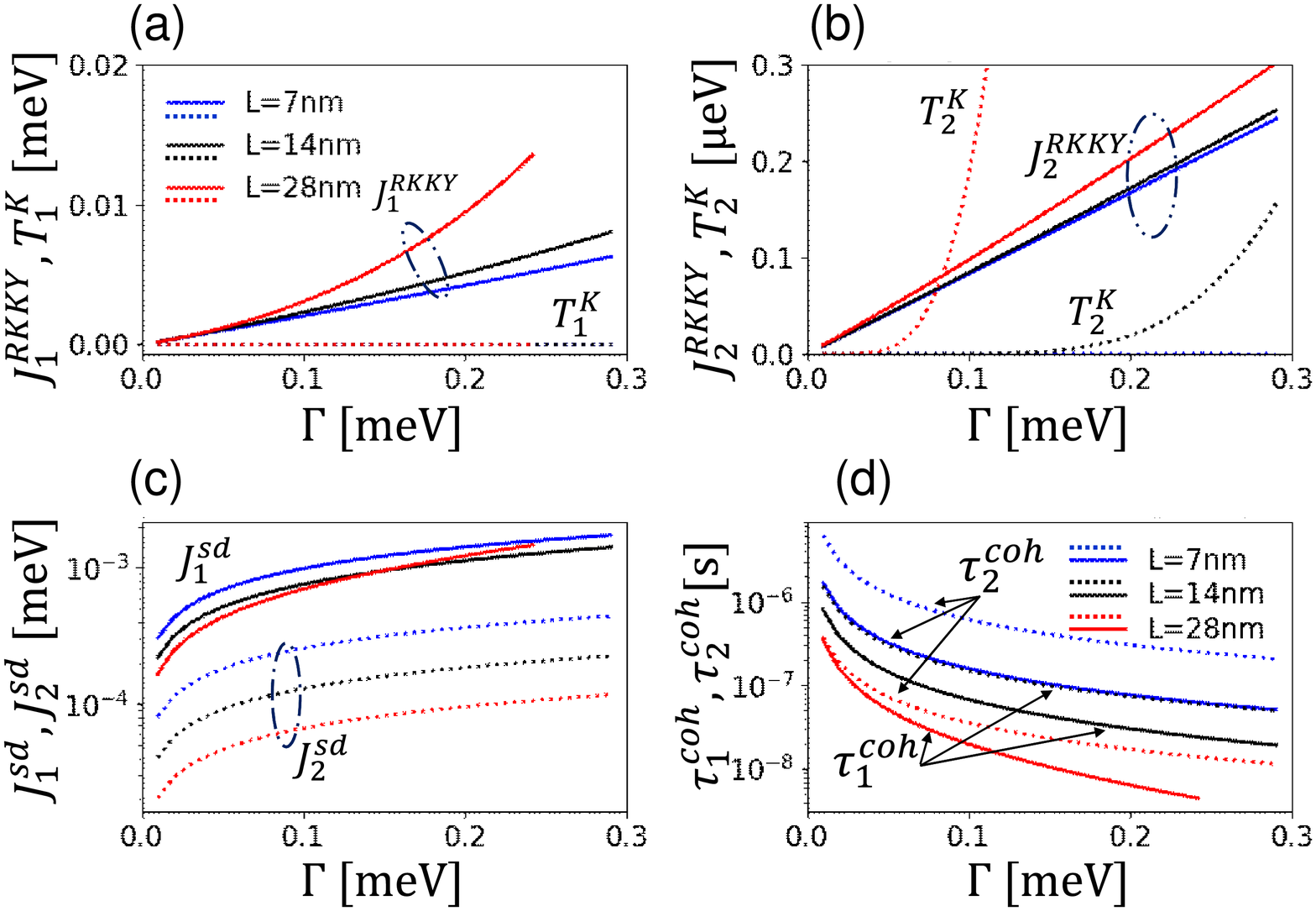}
\includegraphics[width=8.0cm]{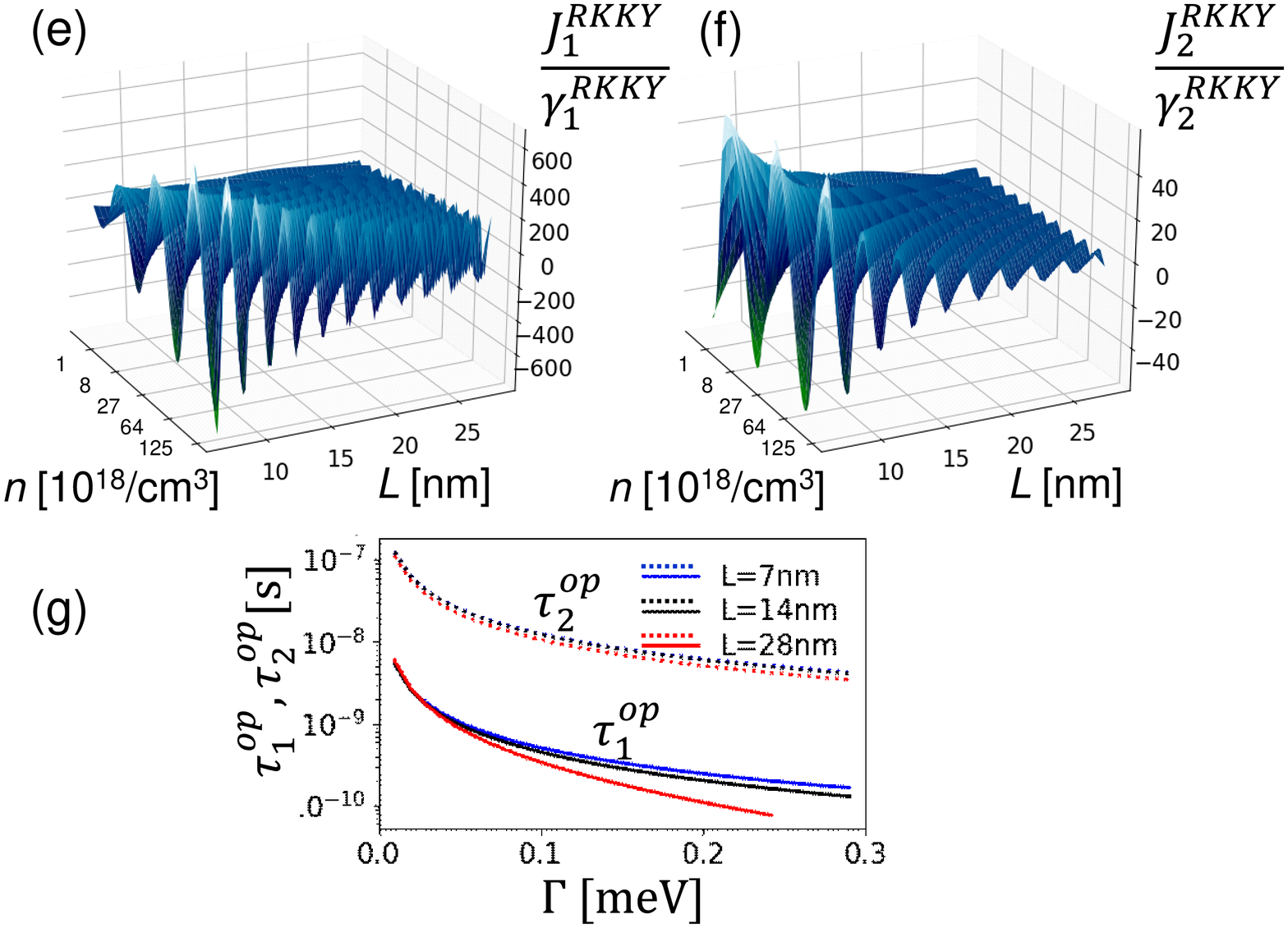}
\end{minipage}
\caption{Various calculated properties of the RKKY interaction.
(a)(b) $J_d^{\rm RKKY}$ and Kondo temperature $T_d^K$ 
as a function of the tunneling strength $\Gamma$ ($d=1,2$). 
(c) Strength of the $s$--$d$ interaction $J_d^{\rm sd}$, 
and (d) coherence times 
at $T=100$~mK. 
In Figs.~a--d, the 1D carrier density is $n_{e1}=0.21$~nm$^{-1}$ , 
and that for 2D is $n_{e2}=0.21^2$~nm$^{-2}$  (9.3 $\times 10^{18}$~cm${}^{-3}$).
(e)(f) Ratio of the RKKY interaction $J^{\rm RKKY}_d$ and
the decoherence rate $\gamma_d$ as a function of the distance 
$L(=W)$ between the two QDs and the carrier density in
the channel $n$ at $T=100$~mK ($n$ is expressed in units per cm$^{-3}$). 
(e) for 1D and (f) for 2D.
(g) Operation times of the $\sqrt{\rm SWAP}$ gate for 1D ($\tau_1^{op}$) and 2D($\tau_2^{op}$)
as a function of $\Gamma$.}
\label{figRKKY}
\end{figure*}

The RKKY interaction is caused by the $s$--$d$ interaction
between the QDs and the channel. 
The magnitude of the $s$--$d$ interaction $J_{\rm sd}$ is 
derived from the tunneling Hamiltonian such that
$
J_{\rm sd}\approx V_{\rm tun}^2 U/(U-E_m)/E_m,
$
where $E_m=E_F-\epsilon_0$~\cite{Coqblin}. 
Thus, we can change $J_{\rm sd}$ by 
controlling $E_F$ through $V_{\rm G}$.
It is convenient to use $z_d\equiv \frac{\Gamma U}{(U-E_m)E_m}$ to express $J_{\rm sd}$ given by
\begin{equation}
J_{\rm sd}= \frac{\Gamma U}{\pi \rho_F(U-E_m)E_m}=\frac{z_d}{\pi \rho_F}.
\label{Jdef}
\end{equation}
In this equation~(\ref{Jdef}), there are restrictions of  
$V_{\rm tun} \ll E_m \ll U-V_{\rm tun}$ and $V_{\rm tun} \ll U/2$~\cite{Coqblin}, 
which lead to $\Gamma \ll \Gamma_{\rm max}\equiv \pi \rho(E_F) U^2/4$.
As $E_m$ decreases ($E_F$ is close to $\epsilon_0$), 
$J_{\rm sd}$ increases, and we take $E_m =2 V_{\rm tun}$ as an example.
One of the advantages of using the transistors is that the carrier density 
can be changed by the gate electrodes $V_G$.
Hereafter, we describe parameters by using the carrier densities $n_{ed}$ 
intended to represent $V_G$ (Table~I).

The 1D and 2D RKKY interactions 
$J_d^{\rm RKKY}$ and the decoherence rate $\gamma_d^{\rm RKKY}$ ($d=1,2$)
are estimated using the formulas of Ref.~\cite{Rikitake}.
They are given by 
$ 
J_d^{\rm RKKY} =\alpha_d \eta_d E_F F_d'(k_F W)$,
and
$\gamma_d^{\rm RKKY} = 4\alpha_d k_B T$,    
where $\alpha_1=m^{*2}J_{\rm sd}^2/(2\pi\hbar^4 k_F^2)$,  
$\alpha_2=m^{*2}J_{\rm sd}^2/32\pi^2\hbar^4$, 
$\eta_1=2$,  
$\eta_2=8/\pi$, and $F_d'(x)$ is a Bessel function (Table~I). 
The coherence time is given by $\tau_{\rm coh}=\hbar/\gamma^{\rm RKKY}_d$.
Although $\gamma_d^{\rm RKKY}$ originally includes Bessel functions,
we use the constant part of $\gamma_d^{\rm RKKY}$ 
to estimate the shortest coherence time~\cite{Rikitake} (see also Appendix~\ref{Idling}). 
Using the $z_d$ defined in equation~(\ref{Jdef}), 
we obtain 
$J_{\rm sd}= \hbar k_F z_1/m^*$ for 1D and $J_{\rm sd}=z_2\hbar^2/m^* $ for 2D, and
\begin{eqnarray}
J_d^{\rm RKKY}&=& \frac{z_d^2 E_F}{\pi} \xi_d F_d'(k_FW), 
\end{eqnarray}
where $\xi_1=1$ and $\xi_2=1/(4\pi^2)$ (Table~I).
The Kondo temperature $T^K$ estimated by
$
T^K 
\approx 
\sqrt{\Gamma U}/2 \exp(\pi \epsilon_0 (\epsilon_0 +U)/[\Gamma U])
$
\cite{Kastner} is rewritten as
\begin{eqnarray}
T_d^K &=& \frac{\sqrt{\Gamma U}}{2} \exp (-\pi/z_d).
\end{eqnarray}

Figures~\ref{figRKKY}a and b show $J_d^{\rm RKKY}$ and $T^K_d$  
as a function of $\Gamma$.
We can see $J_1^{\rm RKKY} > T_1^{\rm K}$ for all $L$s, but 
the region of $J_2^{\rm RKKY} > T_2^{\rm K}$ becomes narrower 
as $L$ increases.
The magnitude of $J_1^{\rm RKKY}$ is much larger than that of $J_2^{\rm RKKY}$, 
reflecting the corresponding magnitudes of $J_d^{sd}$ in Fig.~\ref{figRKKY}~c.
For example, for $\Gamma=0.2$~meV, 
the magnitude of $J_1^{\rm RKKY}$ of $L=28$ nm is approximately 0.01~meV ($\sim$116~mK) and 
that of $J_2^{\rm RKKY}$ of $L=14$~nm is approximately 0.2~$\mu$eV ($\sim$2.32~mK).
Thus, the 1D case is better than the 2D case.
It is also seen that larger $L$ enables larger $J_d^{\rm RKKY}$, because 
$J_d^{\rm RKKY}$ is proportionate to $E_F$.
However, as Fig.~\ref{figRKKY}~d shows, larger $L$ induces shorter coherence time.
Because $J_d^{\rm RKKY}$ is a function of $k_FW$, the relative magnitude of 
$L(=W)$ dependence changes depending on $L$ (see Figs.~\ref{figRKKY}e and f).

In the Heisenberg coupling, $\sqrt{\rm SWAP}$ is the basic element of the 
operation, which requires a time $\tau^{\rm op}$ determined by $J^{\rm RKKY} \tau^{\rm op}=\hbar \pi/2$.
The number of possible operations is estimated using the number of possible operations during the coherence time, 
given by
\begin{equation}
\frac{\tau^{\rm coh}_d}{\tau^{\rm op}_d}\equiv 
\frac{2J^{\rm RKKY}_d}{\pi \gamma^{\rm RKKY}_d}
=  \frac{\eta_d E_F}{2\pi k_B T }  F_d'(k_F W).
\label{decoherenceRKKYratio}
\end{equation}
Because $E_F=\hbar^2 k_F^2/(2m^*)$ and $k_F$ is expressed by the 
density $n_{ed}$ (Table~I), this equation indicates that the ratio is 
determined by $T$, $m^*$, $n_{ed}$, and $W$.
Figures~\ref{figRKKY}e (1D) and f (2D)  show the ratios $ J^{\rm RKKY}_d/\gamma^{\rm RKKY}_d$
as functions of the density $n_{ed}$ and the distance $W(=L)$ 
between the two qubits. 
The oscillations in the figure originate from the Bessel functions.
As the device size $W$ decreases, the number of possible operations 
increases.
Figure~\ref{figRKKY}g shows the time of $\sqrt{\rm SWAP}$.
In addition, for smaller $W(=L)$, 
the 1D cases appear preferable because it can be seen 
that a number in the order of 10${}^2$ operations are possible.
The $s$--$d$ interaction is affected by the magnetic fields~\cite{supplement};
therefore, the RKKY interaction is also affected by $B_z$.
However, we assume $B_z <$ 1 T ($\sim$0.11~meV), 
which means $B_z \ll E_F(\sim$200~meV) and we can neglect 
the effect of $B_z$ in the form of the RKKY interaction.

\subsection{Crosstalk}
As shown in Fig.~1, each LCL affects the neighboring qubits 
(referred to as the crosstalk problem).
This problem can be mitigated by
changing the direction of the neighboring current lines~\cite{Li}.
The detailed analysis and condition are presented in the Appendix~\ref{crosstalk}.

\section{The variation of the size of QDs}
Because the size of the QDs is less than 28nm, the variation of the size of the QDs is unavoidable.
In this section, we consider the effects of the variations of the QDs on the device operations.
When the energy-levels of the QDs are given by $\epsilon_n=\sum_{l=x,y,z}\pi^2\hbar^2 (n_l+1)^2/(2m^*L_{\rm Q}^2)$ 
where $n_x,n_y,n_z=0,1,2...$,
the effect of the variation $L_{\rm Q} \rightarrow  L_{\rm Q}+\Delta L$ induces the 
variation of the energy-levels given by
\begin{equation}
\Delta \epsilon_n=15.5\frac{\Delta L}{L_{\rm Q}}  \sum_{l=x,y,z}  (n_l+1)^2 \ {\rm meV}
\end{equation}
for $L_{\rm Q}=10$ nm (we use $\hbar^2/(2m)=a_0^2 Ry$ and $m^*/m_0=0.5$).
For example, when $\Delta L/L_{\rm Q}=0.1$ that corresponds to 1 nm variations, 
it is possible that the ground state 
variates around $\Delta \epsilon_0 \sim 4.65$~meV. 
As shown in Appendix~\ref{parameters}, the on-site Coulomb energy is given by $U\sim 46.4$meV, 
it is expected that the operations are carried out by adjusting the Fermi energies between the 
nearest two QDs in the range of $U$.
That is, the variations of the size of the QDs are mitigated by controlling the appropriate Fermi levels of 
the channels.
According to the variations of the sizes of the QDs, 
the magnitude of the $J^{\rm RKKY}$ also changes depending on the tunneling
coupling between the QDs and channel.
The time of the two qubit operations should be adjusted depending on 
the individual couplings.
In this process, the appropriate Fermi energies are registered in some digital memory circuits.

The insertion of the excess electrons into each QDs is 
carried by applying voltages between two channels. 
Because each channel is connected to different electrode, 
the transport properties of each QD between the neighboring channels 
can be detected in the same way as the conventional measurement of single-electron devices.

\begin{figure}
\centering
\includegraphics[width=7.5cm]{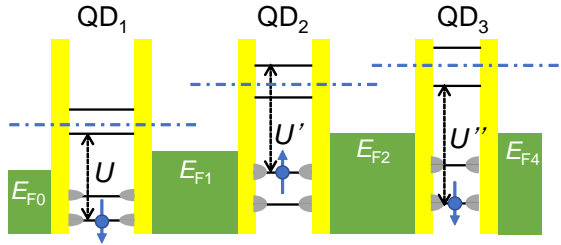}
\caption{The energy diagrams when there are variations of the size of the QDs.
As the size of the QD becomes smaller, the on-site Coulomb energy increases.
The dashed lines show the energy where the spin-filter effects can be detected.
When there are variations, the qubit states are inferred by measuring the channel 
currents with corresponding different source-drain voltages.}
\label{figVAR}
\end{figure}

The channel currents reflect the spin state of neighboring two QDs.
The width of the Zeeman splitting ($\lesssim$ 1~meV) is smaller than the $\Delta \epsilon_0$ 
for $L_{\rm Q}=10$ nm and the energy potentials of the electrodes which enable the spin-filter effect 
are different depending on the channels (Fig.~\ref{figVAR}).
However, because the source-drain current can be measured independently 
by the channel, we can detect the spin-filtered channel current by changing the 
potentials of source and drains in the measurement phase.
As seen from Eq.~(\ref{conductance}), 
the enhancement of the resonant tunneling comes from the energy terms to the 4th power and 
is expected to appear in spite of the existence of the variations.    
The detail analysis of the robustness to the variation is the future problem.

Regarding the corner effect of the FinFETs~\cite{Lansbergen}, because 
we are targeting the 2D electron gas state which is realized 
under relatively larger gate voltage region, 
the effect of the localized state at the corner is considered to be low.
The detail analysis requires TCAD simulations which are beyond the scope of this paper 
and a future problem.

\section{Discussion}
In this paper, we have discussed the conductance of the FinFET devices. 
However, the current--voltage characteristics of the wide ranges of $V_D$ and $V_G$ 
are required to design a large circuit.
The nonlinear current--voltage characteristic is also the origin of the amplifying mechanism 
of the transistors. 
This is a future problem.

In the previous sections, the quantum computations were 
described such that the qubit-- operations 
are carried out by changing the magnitudes of the RKKY interactions.
Instead, the always-on method~\cite{Benjamin} might be suitable for our system,
because in this method, $J^{\rm RKKY}$ is constant 
and the Zeeman energies $\Delta_z$ are adjusted with pulses. 
This method also requires high-frequency control of $B_x$.
Thus, to realize general quantum operations, 
higher-frequency circuits are required.

As mentioned above, the quantum annealer is also a candidate 
device because the high-frequency switching on/off of the 
local fields is not always necessary.
The changing values of RKKY interactions 
depending on gate bias are also suitable for the QAM, 
because the interaction between qubits corresponds 
to the input data of the various combinatorial problems.
It is noted that the present interaction between the qubits 
has Heisenberg form, whereas the conventional QAMs exhibit Ising interactions.
The practical application of the Heisenberg type will be studied in the future~\cite{Cubitt,Xia}.

The dilution refrigerator restricts the power consumption of the chip to the mW order at most.
Assuming a current density of 3 $\times 10^8$~A~cm$^{-2}$  NiSi nanowires~\cite{Wu}
and a resistivity of  10~$\mu \Omega$~cm, 
the power consumption of a wire with an area of 28~nm $\times$ 56~nm and a length of 300~nm 
is given by 1.72 $\times 10^{-10}$~W (the thinnest wire is usually assigned  
only at the lowest layer, referred to as the `M1' layer). 
We can implement approximately 5.8 $\times 10^{6}$ wires in the chip. 
If the thinnest wires are used as connections between the qubits, 
the length of the wire is $L$ and we can afford to use 5.8 $\times 10^{6}$ qubits in a single chip.

\appendix
\section{Equations for estimating parameters}\label{parameters}
Physical parameters are calculated based on basic equations as following.
The Fermi energy 
$E_F=\hbar^2 k_F^2/(2m^*)$ 
of the 1D and 2D  
are given by
\begin{eqnarray}
E_{F1}&=& a_0^2 Ry (\pi n_{e1})^2 (m_0/m^*)\approx 
0.376 n_{e1}^2 (m_0/m^*) {\rm eV}, \nnm \\
E_{F2}&=& a_0^2 Ry (2\pi n_{e2})(m^*/m_0)\approx 
0.239n_{e2} (m_0/m^*) {\rm eV},  \nnm
\end{eqnarray}
where $Ry=13.606$ eV (Rydberg constant), $a_0=0.0529$ nm (Bohr radius), 
and $m_0=9.109\times 10^{-31}$ kg is the electron mass.
The Si effective mass $m^*$ is given by $m^*/m_0=0.2$ for the electrons 
and $m^*/m_0=0.5$ for the holes.
For the density of 10${}^{15}$ cm${}^{-3}$ and 10${}^{18}$ cm${}^{-3}$, 
we have $n_{e1}=0.01$ nm${}^{-1}$ and $n_{e1}=0.1$ nm${}^{-1}$, respectively.

The charging energy is estimated by $U\approx e^2/(2C)$.
where both sides of the capacitance of the QD to the two channels are considered. 
With $C=2 \epsilon_{\rm si} w_d /L_{\rm QD}^2$ 
assuming a cube QD of the size 
$L_{\rm QD}=L/2$, and the thickness of the tunneling barrier $w_d$
($\epsilon_{\rm Si}$ is the dielectric constant of silicon),
we have $U\sim 46.4$ meV for $L=W=10$ nm and $w_d=1$ nm devices.
Note that the gate capacitance changes depending on the $V_G$ and around 1 aF at $V=1V$ in 
Ref.~\cite{Fossum}, which corresponds to 80 meV.


\section{Linear response theory}\label{LRT}
As a typical example, 
we calculate two QDs with three current lines. 
The Hamiltonian of the QDs and the channel is 
given by the tunneling Hamiltonian:
\begin{eqnarray}
H_0&=&E_2 d_{2s}^\dagger d_{2s} +E_4 d_{4s}^\dagger d_{4s}  
+\sum_{i=1,3,5}\sum_{k_i,s} E_{k_i} c_{k_is}^\dagger c_{k_is}   \nnm\\
&+&\sum_{k_1}[V_{k_1} c_{k_1s}^\dagger d_{2s}+V_{k_1}^* d_{2s}^\dagger c_{k_1s} ] \nnm \\
&+&\sum_{k_3} 
[V_{k_3} c_{k_3s}^\dagger (d_{2s}+d_{4s})+V_{k_3}^* (d_{2s}^\dagger+d_{4s}^\dagger)  c_{k_3s} ] \nnm \\
&+&\sum_{k_5}[V_{k_5} c_{k_5s}^\dagger d_{4s}+V_{k_5}^* d_{4s}^\dagger c_{k_5s} ], 
\label{Hamiltonian}
\end{eqnarray}
where the channels are numbered as 1, 3, and 5, and the two QDs are numbered as 2 and 4.
$d_{is}$ and $c_{k,s}$ are the annihilation operators of the QD $i$ 
and the conducting electrons in the channel, respectively.
The qubit states are detected by the channel currents.
The conductance of the channel is calculated using the Kubo formula~\cite{Kubo}.
From Ohm's law, under the electric field $E_y$, 
the current density in the $y$-direction is given by
\begin{equation}
\langle j_y \rangle = \lim_{\omega \rightarrow 0}
g_{yy}(\omega) E_y e^{-i\omega t},
\label{kubo0}
\end{equation}
where the conductance $g_{yy}(\omega)$ is 
calculated from the Kubo formula~\cite{Kubo} given by
\begin{eqnarray}
g_{yy} (\omega) &=&
-\frac{1}{i\omega} [\Phi_{yy}^{\rm R} (\omega)-\Phi_{yy}^{\rm R} (0)], \\
\Phi_{yy}^{\rm R} [t] &=&-\frac{i}{\hbar V}\theta(t) 
\la J_y(t)J_y(0) -J_y(0)J_y(t)\ra. 
\label{kubo1}
\end{eqnarray}
The current operator $J^i_y$ of the $i$th channel is given by
\begin{equation}
J^i_y=(e\hbar/(m^*L))\sum_{k_i} k_i c_{k_is}^\dagger c_{k_is},
\end{equation}
where $L$ is the channel length and the summation of $k_i$ is
carried out over the channel. 
From the current density $j_y$ in equation~(\ref{kubo0}), 
the conventional conductance is given by $G=Vg_{yy}$ ($V$ is a volume).
\begin{eqnarray}
\lefteqn{g_{yy}
\!=\! \frac{2e^2}{h}  k_d \Biggl\{
    \frac{|V_{k_1}|^4(e_4^2\!+\! s_{31}^2)^2}{[(e_1e_4 \!-\! s_{31}^2)^2\!+\! e_4^2\Gamma_1^2]^2}
  + \frac{|V_{k_5}|^4(e_3^2\!+\! s_{35}^2)^2}{[(e_3e_6 \!-\! s_{35}^2)^2\!+\! e_3^2\Gamma_5^2]^2} }\nnm \\
\!&\!+\!&\! \frac{|V_{k_F}|^4 [(e_2+s_{33})^2+ (e_5+s_{33})^2)]^2 }
{\left[ ( e_2e_5\!-\!s_{33}^2)^2 \!+\!\Gamma_3^2 (e_2\!+\!e_5\!+\!2s_{33})^2 \right]^2 }  
\nnm \\ 
\!&\!+\!&\! \frac{2|V_{k_1}|^2|V_{k_3}|^2 (e_4(e_2+s_{33})+s_{31}(e_5+s_{33}))^2}{
[ (e_1e_4 \!-\! s_{31}^2)^2\!+\! e_4^2\Gamma_1^2][( e_2e_5\!-\! s_{33}^2)^2\!+\!\Gamma_3^2 (e_2\!+\! e_5\!+\!2s_{33})^2]}\nnm \\
\!&\!+\!&\! \frac{2|V_{k_1}|^2|V_{k_5}|^2[s_{35}e_4\!+\! s_{31}e_3]^2}{ 
[(e_1e_4 \!-\! s_{31}^2)^2\!+\! e_4^2\Gamma_1^2]
[(e_6e_3 \!-\! s_{35}^2)^2\!+\! e_3^2\Gamma_5^2]} 
 \nnm \\
\!&\!+\!&\! \frac{2|V_{k_3}|^2|V_{k_5}|^2[s_{35}(e_2\! +\! s_{33})\!+\! e_3(e_5\!+\! s_{33})]^2 }{ 
\left[( e_2e_5\!-\! s_{33}^2)^2 \!+\!\Gamma_3^2 (e_2\!+\! e_5\!+\! 2s_{33})^2\right]
[(e_6e_3 \!-\! s_{35}^2)^2\!+\! e_3^2\Gamma_5^2]} 
\Biggr\},
\nnm \\
\label{conductance}
\end{eqnarray}
where $k_1=1$ and $k_2= \pi n_{e2} W^2$. 
\begin{eqnarray}
e_1&=& E_{k_{1}s}-E_2-\Sigma_1(E_{k_1})-\Sigma_3 (E_{k_1}),  \\ 
e_2&=& E_{k_{3}s}-E_2
-\Sigma_1(E_{k_3}) -\Sigma_3(E_{k_3}), \\ 
e_3&=& E_{k_{5}s}-E_2 -\Sigma_1(E_{k_5}) -\Sigma_3(E_{k_5} ), \\
e_4&=& E_{k_{1}s}-E_4-\Sigma_5(E_{k_1})-\Sigma_3(E_{k_1}), \\
e_5&=& E_{k_{3}s}-E_4-\Sigma_5(E_{k_3})-\Sigma_3(E_{k_3}), \\
e_6&=&E_{k_{5}s}-E_4-\Sigma_5(E_{k_5})-\Sigma_3(E_{k_5}),
\end{eqnarray}
where 
\begin{equation}
\Sigma_i (\omega) \equiv \sum_{k_j} \frac{|V_{k_j}|^2}{\omega-E_{k_j}}.
\end{equation}
Detailed derivation is given in the supplementary information.

\section{Idling mode}\label{Idling}
\begin{figure}
\centering
\includegraphics[width=8.0cm]{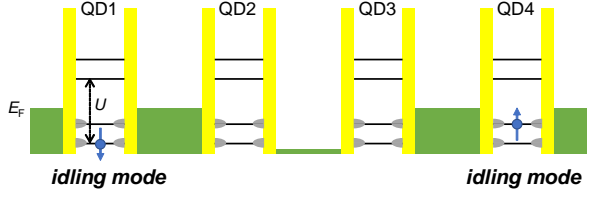}
\caption{Idling mode of the proposed spin qubits.
To maintain charges in QDs, $V_G \neq 0$ is necessary.
When $V_G \neq 0$, neighboring qubits interact via RKKY interactions.
Thus, the independent qubit states (QD1 and QD4) require an extra two QDs (QD2 and QD3) between them. }
\label{fig_idle}
\end{figure}

When $V_G=0$, the Fermi energy is below the energy level of the QDs, 
and the excess electrons leave the QDs. 
Thus, to preserve the qubits, finite $V_G$ is necessary.
This means that this system is a volatile memory. 
Because at present it is difficult to maintain the spin-qubit state 
for more than an hour, this volatile mechanism is sufficient.
When $V_G \neq 0$ and $V_S=V_D$, neighboring qubits exhibit RKKY interactions.
Thus, this system shows an always-on interaction qubit system.
The independent qubit state requires an extra two QDs between them
as shown in Fig.~\ref{fig_idle}.

\section{Noise}\label{Noise}
The shot noise is given by
$S_q=\la \Delta I_q^2 \ra/\Delta f= 2qI=2q g_{yy} V_D$.
For $\Delta g_q=\Delta I_q /V_D$ and $g'=g_{yy} R_K$, we have
\begin{eqnarray}
\Delta g_q' &=& \sqrt{ S_q \Delta f}R_K/V_D =\sqrt{ 2q R_K g' \Delta f/V_D}, 
\end{eqnarray}
where $R_K=h/e^2=25.812$ k$\Omega$ (von Klitzing constant).
The thermal noise is given by $S_T=\la \Delta I_T^2 \ra/\Delta f =4kT g_{yy}$.
For $\Delta g_T=\Delta I_T /V_D$, we have
\begin{eqnarray}
\Delta g_T' &=& \sqrt{ S_T \Delta f}R_K/V_D =\sqrt{ 4kT R_K g' \Delta f /V_D^2}.
\end{eqnarray}
When $\Delta f $ is in the order of 10$^{12}$~s$^{-1}$, $T=$ 100~mK and $V_D=1$~V, 
we have
$
\Delta g_q' = 0.0909 \sqrt{ g' \Delta f/V_D} 
$,
and
$
\Delta g_T' = 3.78\times 10^{-4} \sqrt{ g' \Delta f/V_D^2} 
$.
Thus, we mainly consider the effect of the shot noise.

\section{Coherence time}\label{coherencetime}
In Ref.~\cite{Rikitake}, 
the coherence time is estimated by $\tau_{\rm coh}=\hbar/\gamma^{\rm RKKY}_d$
using the definition of $\gamma^{\rm RKKY}_d$ in Table~II.
The second terms in $G_1'(x)$ and $G_2'(x)$ suppress the relaxation 
between the singlet--triplet transitions, and extend the coherence time.
To estimate the decoherence strictly,
we take $G_1'(x)=G_2'(x)=1$, similarly to in Ref.~\cite{Rikitake} in the text.
\begin{table}
\caption{
$\gamma^{\rm RKKY}$ expressed for 1D ($d=1$) 
and 2D ($d=2$).
$G_d'(k_F W)$s consist of the Bessel functions.
}
\label{table:G}
\begin{center}
\begin{tabular}{l|l|l}\hline
d& $\gamma^{\rm RKKY}_d$ 
 & $G_d'(x)$ \\ \hline
1 
& $\frac{2 z_1^2 k_B T}{\pi}G_1'(k_F W)$
& $G_1'(x)=[1-\cos (2x)]/2$  
 \\ \hline
2  
& $\frac{z_2^2 k_B T}{8\pi^2} G_2'(k_F W)$
& $G_2'(x)=1-J_0^2 (x)$   
 \\ \hline
\end{tabular}
\end{center}
\end{table}

\section{Analysis of crosstalk}\label{crosstalk}
Suppose that there are $N+1$ current lines in parallel. 
The magnetic fields $h_i (i=0,...,N)$ 
estimated by Amp\'ere's law are given by
\begin{eqnarray}
h_0 &=& \frac{1}{2\pi r}[I_0 -pI_1], \   
h_N = (-)^N\frac{1}{2\pi r}[I_N  -pI_{N-1}],   \nnm \\
h_i &=& -\frac{1}{2\pi r}[I_i -p(I_{i-1}+I_{i+1})],\  {\rm for }\ 0<i<N \nnm 
\end{eqnarray}
where $r$ is the distance between the qubits and the current lines, 
$L$ is the distance between the current lines, and $p\equiv r/\sqrt{r^2+L^2}$.
When only the magnetic field of the $n$th qubit is switched on 
while those of the other qubits are switched off, 
the corresponding condition 
$h_1=...=h_{n-1}=h_{n+1}=...=h_N=0$
leads to:
\begin{eqnarray}
I_0&=&  pI_1, \ I_N = pI_{N-1}, \nnm \\
I_i&=&  p(I_{i-1}+I_{i+1}),\   (i\neq n) \nnm \\
h_n  &=& \frac{1}{2\pi r} [I_n -p(I_{n-1} +I_{n+1})]. \nnm  
\end{eqnarray}
Let us consider a case of switching on the $n=3$ qubit out of the six qubits ($N=5$); 
we have
\begin{eqnarray}
I_2&=&\frac{p(1-p^2)}{1-2p^2}I_3,  \
I_4=  \frac{p}{1-p^2} I_3, \nnm \\
I_1&=&\frac{p}{1-p^2} I_2, \ 
I_5=  pI_4, \
I_0=  pI_1.
\end{eqnarray}
The magnetic field to control the third qubit is given by
\begin{eqnarray}
h_3 &=&\left[ 1 -\frac{p^2(1-p^2)}{1-2p^2} -\frac{p^2}{1-p^2}\right] I_3. 
\end{eqnarray}
From this simple analysis, we obtain the condition of the crosstalk problem 
given by $n p^2 \neq 1$ (n = 1, 2, ...), which equals $L\neq \sqrt{n-1}r$.

\section{Fidelity}\label{fidelity}
We assume a Gaussian distribution of conductance.
The conductance $g_{yy}$ is a function of $E_{SR}$ and $E_{SL}$, 
with variation $\Delta g_q$ caused by the noise discussed above.
Thus, when we consider the probabilistic distribution regarding $g_{yy}$,
the conductance $g_{yy}$ is considered to have maximum probability at $g=g_{yy}$ 
and a distribution at around $g_{yy}$ proportionate to 
\begin{equation}
P(g)_{g_{yy}}=\frac{1}{\sqrt{2\pi\Delta g^2_q }}\exp \left\{ -\frac{(g-g_{yy})^2}{2\Delta g^2_q}                   
\right\}.
\end{equation}
The spin direction is determined by comparing the conductance 
with the reference conductance $g_{\uparrow=\downarrow}$.
As $|E_{S\uparrow}-E_{S\downarrow}|$ decreases,
the overlap between $P_{g_{\uparrow\neq\downarrow}}$
and $P_{g_{\uparrow =\downarrow}}$ increases.
Thus, we define the fidelity of the measurement by
\begin{equation}
F_{g_{\uparrow\neq\downarrow}}
=\int {\rm max}\{ P(g)_{g_{\uparrow\neq\downarrow}}
-P(g)_{g_{\uparrow=\downarrow}}, 0 \} dg.
\end{equation}

\subsection*{SUPPLEMENTAL MATERIAL}
See supplemental material for the complete derivation process of the equations.

\begin{acknowledgements}
We acknowledge useful discussions with Takahiro Mori and Hiroshi Fuketa. 
\end{acknowledgements}

\subsection*{DATA AVAILABILITY}
The data that supports the findings of this study are available within the article.



\pagebreak

\onecolumngrid
\begin{center}
  \textbf{\large Compact spin qubits using the common gate structure of fin field-effect transistors\\Supplementary Material}\\[.2cm]
  Tetsufumi Tanamoto,$^{1,*}$ Keiji Ono$^{2}$\\[.1cm]
  {\itshape ${}^1$Department of Information and Electronic Engineering, Teikyo University,
Toyosatodai, Utsunomiya  320-8511, Japan\\
  ${}^2$Advanced device laboratory, RIKEN, Wako-shi, Saitama 351-0198, Japan}
\end{center}

\setcounter{equation}{0}
\setcounter{figure}{0}
\setcounter{table}{0}
\setcounter{page}{1}
\renewcommand{\theequation}{S\arabic{equation}}
\renewcommand{\thefigure}{S\arabic{figure}}
\renewcommand{\bibnumfmt}[1]{[S#1]}
\renewcommand{\citenumfont}[1]{S#1}

\section*{Eigenvalue problem of the resonant-level model}
Before deriving the conductance formula for the two QDs 
coupled with three current channels, we solve the eigenvalue problem 
of the coupled systme of the two QDs and three channels.
This is because, in order to calculate the correlation function of equation (11) in the main text, 
we have to solve the eigenvalue problem of the Hamiltonian (14) in the main text.

As shown in the Fig.7, $i=2$ and $i=4$ indicate the two QDs, and 
$i=1$, $i=3$ and $i=5$ shows the current channel lines. 
Because it is assumed that the spin-flips and decoherence mechanism are neglected, 
we apply the standard method in the range of the resonant-level model following Ref.~\cite{Mahan}.
Then new operators $\alpha_{ks}$ which diagonalize the Hamiltonian 
such as 
$
H=\sum_{i=1,3,5} \sum_{k_i}E_{k_i} \alpha_{k_i}^\dagger  \alpha_{k_i}
$
are introduced:
\begin{eqnarray}
d_{2s} &=& 
  \sum_{k_1s} \nu^{(21)}_{k_1s} \alpha_{k_1s}
 +\sum_{k_3s} \nu^{(23)}_{k_3s} \alpha_{k_3s} 
 +\sum_{k_5s} \nu^{(25)}_{k_5s} \alpha_{k_5s}, \\
d_{4s} &=&
  \sum_{k_1s} \nu^{(41)}_{k_1s} \alpha_{k_1s} 
 +\sum_{k_3s} \nu^{(43)}_{k_3s} \alpha_{k_3s} 
 +\sum_{k_5s} \nu^{(45)}_{k_5s} \alpha_{k_5s},\\
c_{k_1s}&=& 
 \sum_{k_1'} \eta^{(11)}_{k_1,k_1'} \alpha_{k'_1s}
+\sum_{k_3'} \eta^{(13)}_{k_1,k_3'} \alpha_{k'_3s}
+\sum_{k_5'} \eta^{(15)}_{k_1,k_5'} \alpha_{k'_5s},\\
c_{k_3s}&=& 
 \sum_{k_1'} \eta^{(31)}_{k_3,k_1'} \alpha_{k'_1s}
+\sum_{k_3'} \eta^{(33)}_{k_3,k_3'} \alpha_{k'_3s}
+\sum_{k_5'} \eta^{(35)}_{k_3,k_5'} \alpha_{k'_5s},\\
c_{k_5s}&=& 
 \sum_{k_1'} \eta^{(51)}_{k_5,k_1'} \alpha_{k'_1s}
+\sum_{k_3'} \eta^{(53)}_{k_5,k_3'} \alpha_{k'_3s}
+\sum_{k_5'} \eta^{(55)}_{k_5,k_5'} \alpha_{k'_5s}.
\end{eqnarray}
The unknown matrix elements $\eta^{(ij)}_{k_i,k'j,s}$, $\nu^{(ij)}_{k_js}$ 
are determined from expressing the commutation relations $[d_{is}, H]$,$[c_{k_is}, H]$ 
by both the new operators  $\alpha_{k_is}$ and the original operators 
$d_{is}$,and $c_{k_is}$.

For example ($i=2,4$),
\begin{eqnarray}
[d_{is}, H]
&=&[
\sum_{j=1,3,5}\sum_{k_{j}} \nu^{(ij)}_{k_{j}s}\alpha_{k_{j}s},
\sum_{l=1,3,5}\sum_{k_l}E_{k_l} \alpha_{k_l}^\dagger  \alpha_{k_l}]  \nnm \\
&=&
\sum_{j=1,3,5} \sum_{k_{j}} \nu^{(ij)}_{k_{j}s} E_{k_j}\alpha_{k_{j}s}.
\end{eqnarray}
On the other hand
\begin{eqnarray}
[d_{is}, H]
&=&[d_{is},
E_i d_{is}^\dagger d_{is} \nonumber \\
&+&\sum_{k_1}[V_{k_1}^* d_{2s}^\dagger c_{k_1s} ] 
+\sum_{k_3}[V_{k_3}^* (d_{2s}^\dagger+d_{4s}^\dagger)  c_{k_3s} ] 
+\sum_{k_5}[V_{k_5}^* d_{4s}^\dagger c_{k_5s} ] ]\nnm \\
&=&
E_i d_{is} 
+ \sum_{k_1}V_{k_{i-1}}^* c_{k_{i-1}s} 
+ \sum_{k_3}V_{k_{i+1}}^* c_{k_{i+1}s}. 
\end{eqnarray}
\begin{figure}
\centering
\includegraphics[width=6.0cm]{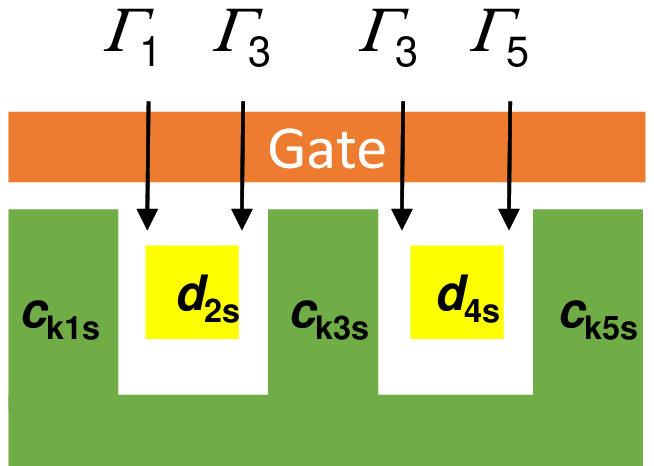}
\caption{Two quantum dots surrounded by three current lines.
$c_{k_is}$ and $d_{is}$ show the annihilation operators.
$\Gamma_i$s are the tunneling couplings derived in this 
supplements.
}
\label{figsup}
\end{figure}

Thus,
\begin{eqnarray}
\lefteqn{
 \sum_{k_{1}s} E_{k_{1}s} \nu^{(21)}_{k_{1}s}\alpha_{k_{1}s}
+\sum_{k_{3}s} E_{k_{3}s} \nu^{(23)}_{k_{3}s}\alpha_{k_{3}s}
+\sum_{k_{5}s} E_{k_{5}s} \nu^{(25)}_{k_{3}s}\alpha_{k_{5}s}
=E_2 (
 \sum_{k_1s} \nu^{(21)}_{k_1s} \alpha_{k_1s}
+\sum_{k_3s} \nu^{(23)}_{k_3s} \alpha_{k_3s}
+\sum_{k_5s} \nu^{(25)}_{k_5s} \alpha_{k_5s})
}\nnm \\
&+& \sum_{k_1}V_{k_{1}}^* (
 \sum_{k_1'} \eta^{(11)}_{k_1,k_1'} \alpha_{k'_1s}
+\sum_{k_3'} \eta^{(13)}_{k_1,k_3'} \alpha_{k'_3s}
+\sum_{k_5'} \eta^{(15)}_{k_1,k_5'} \alpha_{k'_5s}) 
+ \sum_{k_3}V_{k_{3}}^* (
 \sum_{k_1'} \eta^{(31)}_{k_3,k_1'} \alpha_{k'_1s}
+\sum_{k_3'} \eta^{(33)}_{k_3,k_3'} \alpha_{k'_3s}
+\sum_{k_5'} \eta^{(35)}_{k_3,k_5'} \alpha_{k'_5s}). \nnm
\end{eqnarray}
From these equations, we have
\begin{eqnarray}
(E_{k_{1}s}-E_2) \nu^{(21)}_{k_{1}s}
&=&  \sum_{k_1'}V_{k_{1}'}^* \eta^{(11)}_{k_1',k_1}+\sum_{k_3'}V_{k_{3}'}^* \eta^{(31)}_{k_3',k_1}, 
\label{eta1} \\ 
(E_{k_{3}s}-E_2) \nu^{(23)}_{k_{3}s}
&=& \sum_{k'_1}V_{k'_{1}}^*  \eta^{(13)}_{k_1',k_3}+ \sum_{k'_3}V_{k'_{3}}^*  \eta^{(33)}_{k_3',k_3},  
\label{eta2}\\ 
(E_{k_{5}s}-E_2) \nu^{(25)}_{k_{5}s}
&=& \sum_{k'_1}V_{k'_{1}}^*  \eta^{(15)}_{k_1',k_5}+ \sum_{k'_3}V_{k'_{3}}^*  \eta^{(35)}_{k_3',k_5}.  
\label{eta3}
\end{eqnarray}
Similarly we have 
\begin{eqnarray}
(E_{k_{1}s}-E_4) \nu^{(41)}_{k_{1}s}&=&
 \sum_{k_5'}V_{k_{5}'}^* \eta^{(51)}_{k_5',k_1}  
+\sum_{k_3'}V_{k_{3}'}^* \eta^{(31)}_{k_3',k_1},  \label{eta4}\\
(E_{k_{3}s}-E_4) \nu^{(43)}_{k_{3}s}
&=&
  \sum_{k_5'}V_{k_{5}'}^* \eta^{(53)}_{k_5',k_3} 
+ \sum_{k_3'}V_{k_{3}'}^* \eta^{(33)}_{k_3',k_3},  \label{eta5}\\
(E_{k_{5}s}-E_4) \nu^{(45)}_{k_{5}s} 
&=&
 \sum_{k_5'}V_{k_{5}'}^*  \eta^{(55)}_{k_5',k_5}  
+\sum_{k_3'}V_{k_{3}'}^*  \eta^{(35)}_{k_3',k_5}.  \label{eta6}
\end{eqnarray}
The $[c_{k_is},H]$ can be calculated ($i=1,3,5$)
\begin{eqnarray}
[c_{k_is},H] &=&
[c_{k_is}, 
   \sum_{i=1,3,5}\sum_{k_i,s} E_{k_i} c_{k_is}^\dagger c_{k_is}  
+\sum_{k_1}[V_{k_1} c_{k_1s}^\dagger d_{2s}+V_{k_1}^* d_{2s}^\dagger c_{k_1s} ] \nnm \\
&+&\sum_{k_3}[V_{k_3} c_{k_3s}^\dagger (d_{2s}+d_{4s})+V_{k_3}^* (d_{2s}^\dagger+d_{4s}^\dagger)  c_{k_3s} ] 
+\sum_{k_5}[V_{k_5} c_{k_5s}^\dagger d_{4s}+V_{k_5}^* d_{4s}^\dagger c_{k_5s} ] ].
\end{eqnarray} 
On the other hand, we have for $i=1,3,5$,
\begin{eqnarray}
[c_{k_i},H] 
&=&
[\sum_{k_1'} \eta^{(i1)}_{k_i,k_1'} \alpha_{k'_1s}
+\sum_{k_3'} \eta^{(i3)}_{k_i,k_3'} \alpha_{k'_3s}
+\sum_{k_5'} \eta^{(i5)}_{k_i,k_5'} \alpha_{k'_5s},
\sum_{i=1,3,5} \sum_{k_i"}E_{k_i"} \alpha_{k_i"}^\dagger  \alpha_{k_i"} ]
\nnm\\
&=&
 [\sum_{k_1'}\eta^{(i1)}_{k_i,k_1'}\alpha_{k'_1s},\sum_{k_i"}E_{k_i"} \alpha_{k_i"}^\dagger  \alpha_{k_i"}]
+[\sum_{k_3'}\eta^{(i3)}_{k_i,k_3'}\alpha_{k'_3s},\sum_{k_i"}E_{k_i"} \alpha_{k_i"}^\dagger  \alpha_{k_i"}]
+[\sum_{k_5'}\eta^{(i5)}_{k_i,k_5'} \alpha_{k'_5s},\sum_{k_i"}E_{k_i"} \alpha_{k_i"}^\dagger  \alpha_{k_i"} ]
\nnm\\
&=&
 \sum_{k_1'}\eta^{(i1)}_{k_i,k_1'}E_{k_1'}\alpha_{k'_1s}
+\sum_{k_3'}\eta^{(i3)}_{k_i,k_3'}E_{k_3'}\alpha_{k'_3s}
+\sum_{k_5'}\eta^{(i5)}_{k_i,k_5'}E_{k_5'}\alpha_{k'_5s}.
\nnm
\end{eqnarray}
For example, the case of $i=1$ shows
\begin{eqnarray}
& & \sum_{k_1's} [E_{k_1}\eta^{(11)}_{k_1,k_1'}+V_{k_1} \nu^{(21)}_{k_1's}]\alpha_{k_1's}
+\sum_{k_3's} [E_{k_1}\eta^{(13)}_{k_1,k_3'}+V_{k_1} \nu^{(23)}_{k_3's}]\alpha_{k_3's}
+\sum_{k_5's} [E_{k_1}\eta^{(15)}_{k_1,k_5'}+V_{k_1} \nu^{(25)}_{k_5's}]\alpha_{k_5's}
\nnm \\
&=&
 \sum_{k_1'}\eta^{(11)}_{k_1,k_1'}E_{k_1'}\alpha_{k'_1s}
+\sum_{k_3'}\eta^{(13)}_{k_1,k_3'}E_{k_3'}\alpha_{k'_3s}
+\sum_{k_5'}\eta^{(15)}_{k_1,k_5'}E_{k_5'}\alpha_{k'_5s}.
\nnm 
\end{eqnarray}
Thus, we get
\begin{eqnarray}
& &[E_{k_1}-E_{k_1'}]\eta^{(11)}_{k_1,k_1'}+V_{k_1}\nu^{(21)}_{k_1's} = 0, \nnm \\
& &[E_{k_1}-E_{k_3'}]\eta^{(13)}_{k_1,k_3'}+V_{k_1}\nu^{(23)}_{k_3's} = 0, \nnm \\
& &[E_{k_1}-E_{k_5'}]\eta^{(15)}_{k_1,k_5'}+V_{k_1}\nu^{(25)}_{k_5's} = 0. \nnm 
\end{eqnarray}
Similarly we have 
\begin{eqnarray}
& &[E_{k_3}-E_{k_1'}]\eta^{(31)}_{k_3,k_1'} 
+V_{k_3} \nu^{(21)}_{k_1's} 
+V_{k_3} \nu^{(41)}_{k_1's} =0, 
\nnm \\
& &[E_{k_3}-E_{k_3'}] \eta^{(33)}_{k_3,k_1'} 
+V_{k_3} \nu^{(23)}_{k_3's} 
+V_{k_3} \nu^{(43)}_{k_3's} =0,
\nnm \\
& &[E_{k_3}-E_{k_5'}]\eta^{(35)}_{k_3,k_5'} 
+V_{k_3} \nu^{(25)}_{k_5's} 
+V_{k_3} \nu^{(45)}_{k_5's} =0,
\nnm \\
& &[E_{k_5}-E_{k_1'}]\eta^{(51)}_{k_5,k_1'}+V_{k_5} \nu^{(41)}_{k_1's}=0,\nnm \\
& &[E_{k_5}-E_{k_3'}]\eta^{(53)}_{k_5,k_3'}+V_{k_5} \nu^{(43)}_{k_3's}=0,\nnm\\
& &[E_{k_5}-E_{k_5'}]\eta^{(55)}_{k_5,k_5'}+V_{k_5} \nu^{(45)}_{k_5's}=0.\nnm
\end{eqnarray} 
Thus, we get
\begin{eqnarray}
& &\eta^{(11)}_{k_1',k_1}=-\frac{V_{k_1'}}{E_{k_1'}-E_{k_1}}\nu^{(21)}_{k_1s} 
+\delta_{k_1}^{k_1'}Z_{k_1}V_{k_1} \nu^{(21)}_{k_1s},\label{eta1} \\
& &\eta^{(13)}_{k_1',k_3}=-\frac{V_{k_1'}}{E_{k_1'}-E_{k_3}}\nu^{(23)}_{k_3s},  \\
& &\eta^{(15)}_{k_1',k_5}=-\frac{V_{k_1'}}{E_{k_1'}-E_{k_5}}\nu^{(25)}_{k_5s},  \\
& &\eta^{(31)}_{k_3',k_1}=-\frac{V_{k_3'}}{E_{k_3'}-E_{k_1}}[\nu^{(21)}_{k_1s}+\nu^{(41)}_{k_1s}],\\
& &\eta^{(33)}_{k_3',k_3}=-\frac{V_{k_3'}}{E_{k_3'}-E_{k_3}}[\nu^{(23)}_{k_3s}+\nu^{(43)}_{k_3s}],
+\delta_{k_3}^{k_3'}Z_{k_3}V_{k_3} [\nu^{(23)}_{k_3s}+\nu^{(43)}_{k_3s}], \\
& &\eta^{(35)}_{k_3',k_5}=-\frac{V_{k_3'}}{E_{k_3'}-E_{k_5}}[\nu^{(25)}_{k_5s}+\nu^{(45)}_{k_5s}], \\
& &\eta^{(51)}_{k_5',k_1}=-\frac{V_{k_5'}}{E_{k_5'}-E_{k_1}} \nu^{(41)}_{k_1s}, \\
& &\eta^{(53)}_{k_5',k_3}=-\frac{V_{k_5'}}{E_{k_5'}-E_{k_3}} \nu^{(43)}_{k_3s}, \\
& &\eta^{(55)}_{k_5',k_5}=-\frac{V_{k_5'}}{E_{k_5'}-E_{k_5}} \nu^{(45)}_{k_5s} 
+\delta_{k_5}^{k_5'}Z_{k_5}V_{k_5} \nu^{(45)}_{k_5s}.  \label{eta9}
\end{eqnarray} 
By substituting these equations into Eqs.(\ref{eta1})-(\ref{eta6}), we have
\begin{eqnarray}
\left[ E_{k_{1}s}-E_2-\Sigma_1(E_{k_1})-\Sigma_3 (E_{k_1})\right] \nu^{(21)}_{k_{1}s}
-\Sigma_3 (E_{k_1})[\nu^{(41)}_{k_1s}]
&=&  Z_{k_1}|V_{k_1}|^2 \nu^{(21)}_{k_1s},
\nnm \\ 
\left[E_{k_{3}s}-E_2
-\Sigma_1(E_{k_3}) -\Sigma_3(E_{k_3})\right]\nu^{(23)}_{k_3s}
-\Sigma_3(E_{k_3}) [\nu^{(43)}_{k_3s}]&=&
Z_{k_3}|V_{k_3}|^2 [\nu^{(23)}_{k_3s}+\nu^{(43)}_{k_3s}],  
\nnm \\ 
\left[ E_{k_{5}s}-E_2 -\Sigma_1(E_{k_5}) -\Sigma_3(E_{k_5} ) \right]\nu^{(25)}_{k_5s}
-\Sigma_3(E_{k_5} )[\nu^{(45)}_{k_5s}] 
&=& 0, 
\nnm \\
\left[E_{k_{1}s}-E_4-\Sigma_5(E_{k_1})-\Sigma_3(E_{k_1})\right]\nu^{(41)}_{k_{1}s}
-\Sigma_3(E_{k_1})[\nu^{(21)}_{k_1s}] 
&=& 0,  
\nnm \\
\left[E_{k_{3}s}-E_4-\Sigma_5(E_{k_3})-\Sigma_3(E_{k_3})\right]\nu^{(43)}_{k_{3}s}
-\Sigma_3(E_{k_3})[\nu^{(23)}_{k_3s}] 
&=& Z_{k_3}|V_{k_3}|^2 [\nu^{(23)}_{k_3s}+\nu^{(43)}_{k_3s}], 
\nnm \\
\left[E_{k_{5}s}-E_4-\Sigma_5(E_{k_5})-\Sigma_3(E_{k_5})\right]\nu^{(45)}_{k_{5}s} 
-\Sigma_3(E_{k_5})[\nu^{(25)}_{k_5s}]  
&=& Z_{k_5}|V_{k_5}|^2 \nu^{(45)}_{k_5s}, 
\nnm
\end{eqnarray}
where
\begin{eqnarray}
\Sigma_1(E_{k})&=&  -\sum_{k_1'}\frac{|V_{k_1'}|^2}{E_{k_1'}-E_{k}}, 
\nnm \\ 
\Sigma_3(E_{k}) &=& -\sum_{k'_3}\frac{|V_{k_3'}|^2}{E_{k_3'}-E_{k}},
\nnm \\ 
\Sigma_5(E_{k}) &=& -\sum_{k_5'}\frac{|V_{k_5'}|^2}{E_{k_5'}-E_{k}}. \nnm
\end{eqnarray}

Next,  the relation $\{d_{i},d_{j}^\dagger \}=\delta_{ij}$ leads to:
\begin{eqnarray}
\{d_{i},d_{j}^\dagger \}&=&
\{
  \sum_{k_1'} \nu^{(i1)}_{k_1's} \alpha_{k_1's}
 +\sum_{k_3'} \nu^{(i3)}_{k_3's} \alpha_{k_3's}
 +\sum_{k_5'} \nu^{(i5)}_{k_5's} \alpha_{k_5's},
  \sum_{k_1"s} \nu^{(j1)*}_{k_1"s} \alpha_{k_1"s}^\dagger
 +\sum_{k_3"s} \nu^{(j3)*}_{k_3"s} \alpha_{k_3"s}^\dagger 
 +\sum_{k_5"s} \nu^{(j5)*}_{k_5"s} \alpha_{k_5"s}^\dagger
\}
\nnm \\
&=&
 \sum_{k_1'} \nu^{(i1)}_{k_1's}\nu^{(j1)*}_{k_1's}
+\sum_{k_3'} \nu^{(i3)}_{k_3's}\nu^{(j3)*}_{k_3's}
+\sum_{k_5'} \nu^{(i5)}_{k_5's}\nu^{(j5)*}_{k_5's}.
\end{eqnarray}
Then, we have 
\begin{eqnarray}
& &
 \sum_{k_1'} |\nu^{(21)}_{k_1's}|^2
+\sum_{k_3'} |\nu^{(23)}_{k_3's}|^2
+\sum_{k_5'} |\nu^{(25)}_{k_5's}|^2=1,
\\
& &
 \sum_{k_1'} |\nu^{(41)}_{k_1's}|^2
+\sum_{k_3'} |\nu^{(43)}_{k_3's}|^2
+\sum_{k_5'} |\nu^{(45)}_{k_5's}|^2=1,
\\
& &
 \sum_{k_1'} \nu^{(21)}_{k_1's}\nu^{(41)*}_{k_1's}
+\sum_{k_3'} \nu^{(23)}_{k_3's}\nu^{(43)*}_{k_3's}
+\sum_{k_5'} \nu^{(25)}_{k_5's}\nu^{(45)*}_{k_5's}=0.
\end{eqnarray}

The $\{d_{k_i},c_{k_j}^\dagger \}$ leads to
\begin{eqnarray}
\{c_{k_i},d_{j}^\dagger \}&=&
\{
 \sum_{k_1'} \eta^{(i1)*}_{k_i,k_1'} \alpha_{k'_1s}
+\sum_{k_3'} \eta^{(i3)*}_{k_i,k_3'} \alpha_{k'_3s}
+\sum_{k_5'} \eta^{(i5)*}_{k_i,k_5'} \alpha_{k'_5s},
  \sum_{k_1"s} \nu^{(j1)}_{k_1"s} \alpha_{k_1"s}^\dagger
 +\sum_{k_3"s} \nu^{(j3)}_{k_3"s} \alpha_{k_3"s}^\dagger 
 +\sum_{k_5"s} \nu^{(j5)}_{k_5"s} \alpha_{k_5"s}^\dagger,
\}
\nnm \\
&=&
 \sum_{k_1'} \eta^{(i1)}_{k_i,k_1'}\nu^{(j1)*}_{k_1's}
+\sum_{k_3'} \eta^{(i3)}_{k_i,k_3'}\nu^{(j3)*}_{k_3's}
+\sum_{k_5'} \eta^{(i5)}_{k_i,k_5'}\nu^{(j5)*}_{k_5's}=0.
\end{eqnarray}
For example, for $(i,j)=(1,2)$, we have
\begin{eqnarray}
0&=&
 \sum_{k_1'} \eta^{(11)}_{k_1,k_1'}\nu^{(21)*}_{k_1's}
+\sum_{k_3'} \eta^{(13)}_{k_1,k_3'}\nu^{(23)*}_{k_3's}
+\sum_{k_5'} \eta^{(15)}_{k_1,k_5'}\nu^{(25)*}_{k_5's}
\nnm \\
&=&
 \sum_{k_1'} (-\frac{V_{k_1}}{E_{k_1}-E_{k_1'}}\nu^{(21)}_{k_1's} 
+\delta_{k_1}^{k_1'}Z_{k_1}V_{k_1} \nu^{(21)}_{k_1s})\nu^{(21)*}_{k_1's}
+\sum_{k_3'} (-\frac{V_{k_1}}{E_{k_1}-E_{k_3'}}\nu^{(23)}_{k_3's})\nu^{(23)*}_{k_3's}
+\sum_{k_5'} (-\frac{V_{k_1}}{E_{k_1}-E_{k_5'}}\nu^{(25)}_{k_5s} )\nu^{(25)*}_{k_5's}
\nnm\\
&=&
 \sum_{k_1'} (-\frac{V_{k_1}}{E_{k_1}-E_{k_1'}}
+\delta_{k_1}^{k_1'}Z_{k_1}V_{k_1})|\nu^{(21)}_{k_1's}|^2
-\sum_{k_3'} \frac{V_{k_1}}{E_{k_1}-E_{k_3'}}|\nu^{(23)}_{k_3's}|^2
-\sum_{k_5'} \frac{V_{k_1}}{E_{k_1}-E_{k_5'}}|\nu^{(25)}_{k_5's}|^2.
\end{eqnarray}
For example, for $(i,j)=(3,2)$, we have
\begin{eqnarray}
0&=&
 \sum_{k_1'} \eta^{(31)}_{k_3,k_1'}\nu^{(21)*}_{k_1's}
+\sum_{k_3'} \eta^{(33)}_{k_3,k_3'}\nu^{(23)*}_{k_3's}
+\sum_{k_5'} \eta^{(35)}_{k_3,k_5'}\nu^{(25)*}_{k_5's}
\nnm \\
&=&
 \sum_{k_1'} (-\frac{V_{k_3}}{E_{k_3}-E_{k_1'}}[\nu^{(21)}_{k_1's}+\nu^{(41)}_{k_1's}])\nu^{(21)*}_{k_1's}
+\sum_{k_3'} (-\frac{V_{k_3}}{E_{k_3}-E_{k_3'}}[\nu^{(23)}_{k_3's}+\nu^{(43)}_{k_3's}]
+\delta_{k_3}^{k_3'}Z_{k_3}V_{k_3} [\nu^{(23)}_{k_3s}+\nu^{(43)}_{k_3s}])\nu^{(23)*}_{k_3's}
\nnm \\
&+&\sum_{k_5'} (-\frac{V_{k_3}}{E_{k_3}-E_{k_5'}}[\nu^{(25)}_{k_5's}+\nu^{(45)}_{k_5's})\nu^{(25)*}_{k_5's}.
\nnm
\end{eqnarray}
%
We can use other commutation relation $\{c_{k_i},c_{k_j''}^\dagger \}=\delta_{k_i,k_j''}$ such as
\begin{eqnarray}
\delta_{k_i,k_j''} &=&\{c_{k_i},c_{k_j''}^\dagger \}\nnm \\
&=&
 \sum_{k_1'} \eta^{(i1)}_{k_i,k_1'} \eta^{(j1)*}_{k_j'',k_1'} 
+\sum_{k_3'} \eta^{(i3)}_{k_i,k_3'} \eta^{(j3)*}_{k_j'',k_3'} 
+\sum_{k_5'} \eta^{(i5)}_{k_i,k_5'} \eta^{(j5)*}_{k_j'',k_5'}.
\end{eqnarray}
For example, for $(i,j)=(1,3)$, we have
\begin{eqnarray}
\lefteqn{
\delta_{k_1,k_3''}=0
=\sum_{k_1'}(-\frac{V_{k_1}}{E_{k_1}-E_{k_1'}}\nu^{(21)}_{k_1's} 
+\delta_{k_1}^{k_1'}Z_{k_1}V_{k_1}  \nu^{(21)}_{k_1s})
(
-\frac{V_{k_3''}^*}{E_{k_3''}-E_{k_1'}}[\nu^{(21)*}_{k_1's}+\nu^{(41)*}_{k_1's}])
}\nnm\\
&+&
\sum_{k_3'}(-\frac{V_{k_1}}{E_{k_1}-E_{k_3'}}\nu^{(23)}_{k_3's})
(-\frac{V_{k_3''}^*}{E_{k_3''}-E_{k_3'}}[\nu^{(23)*}_{k_3's}+\nu^{(43)*}_{k_3's}]
+\delta_{k_3'}^{k_3''}Z_{k_3'}V_{k_3'}^* [\nu^{(23)*}_{k_3's}+\nu^{(43)*}_{k_3's}])
\nnm\\
&+&\sum_{k_5'}(-\frac{V_{k_1}}{E_{k_1}-E_{k_5'}}\nu^{(25)}_{k_5's})
(-\frac{V_{k_3''}^*}{E_{k_3''}-E_{k_5'}}[\nu^{(25)*}_{k_5's}+\nu^{(45)*}_{k_5's}])
\nnm\\
&=&V_{k_1}V_{k_3''}^*(
\sum_{k_1'}(
\frac{1}{E_{k_3''}-E_{k_1'}}(\frac{1}{E_{k_1}-E_{k_1'}}-\frac{1}{E_{k_3''}-E_{k_1'}})
 \nu^{(21)}_{k_1's}[\nu^{(21)*}_{k_1's}+\nu^{(41)*}_{k_1's}]
-\delta_{k_1}^{k_1'}Z_{k_1}\nu^{(21)}_{k_1s}\frac{1}{E_{k_3''}-E_{k_1'}}
[\nu^{(21)*}_{k_1's}+\nu^{(41)*}_{k_1's}]
)
\nnm\\
&+&
\sum_{k_3'}
(\frac{1}{E_{k_3''}-E_{k_3'}}(\frac{1}{E_{k_1}-E_{k_3'}}-\frac{1}{E_{k_3''}-E_{k_3'}})
\nu^{(23)}_{k_3's}[\nu^{(23)*}_{k_3's}+\nu^{(43)*}_{k_3's}]
+\frac{1}{E_{k_3''}-E_{k_1}}\nu^{(23)}_{k_3's}\delta_{k_3'}^{k_3''}Z_{k_3'}[\nu^{(23)*}_{k_3's}+\nu^{(43)*}_{k_3's}])
\nnm\\
&+&
\sum_{k_5'}\frac{1}{E_{k_3''}-E_{k_5'}}(\frac{1}{E_{k_1}-E_{k_5'}}-\frac{1}{E_{k_3''}-E_{k_5'}})
\nu^{(25)}_{k_5's}[\nu^{(25)*}_{k_5's}+\nu^{(45)*}_{k_5's}]
).
\nnm
\end{eqnarray}
For $(i,j)=(1,1)$ we have
\begin{eqnarray}
\delta_{k_1,k_1''} 
&=&\sum_{k_1'}V_{k_1}V_{k_1''^*} |\nu^{(21)}_{k_1's}|^2(
 \frac{L^2}{2u_k^2} \delta_{k_1,k_1"}\delta_{k_1',k_1"}
+\delta_{k_1}^{k_1'}\delta_{k_1''}^{k_1'}Z_{k_1}Z_{k_1''})
\nnm\\
&+&
 \frac{1}{E_{k_1''}-E_{k_1}}  \{
 \sum_{k_1'}V_{k_1}V_{k_1''^*} |\nu^{(21)}_{k_1's}|^2(
 \frac{1}{E_{k_1}-E_{k_1'}}-\frac{1}{E_{k"_1}-E_{k_1'}}
+\delta_{k_1''}^{k_1'}Z_{k_1''}
-\delta_{k_1'}^{k_1}Z_{k_1})
\nnm\\
&+&
\sum_{k_3'}V_{k_1}V_{k_1''}^*|\nu^{(23)}_{k_3's}|^2
(\frac{1}{E_{k_1}-E_{k_3'}}-\frac{1}{E_{k_1''}-E_{k_3'}})
+
\sum_{k_5'}V_{k_1}V_{k_1''}^*|\nu^{(25)}_{k_5's}|^2
(\frac{1}{E_{k_1}-E_{k_5'}}-\frac{1}{E_{k_1''}-E_{k_5'}})
\}
\nnm\\
&=& |V_{k_1}|^2 
|\nu^{(21)}_{k_1s}|^2(
 Z_{k_1}^2+\frac{L^2}{2u_k^2} )\delta_{k_1,k_1''}.
\end{eqnarray}

In these derivations, we have used Poincare's relation given by
\begin{eqnarray}
P\frac{1}{E_{k_L}-E_{k_L''}}\frac{1}{E_{k_L'}-E_{k_L''}}
=
P\frac{1}{E_{k_L}-E_{k_L'}}
\left( 
\frac{1}{E_{k_L''}-E_{k_L}}-\frac{1}{E_{k_L''}-E_{k_L'}}
\right)
+ \frac{L^2}{2u_k^2} \delta_{k_L,k_L"}\delta_{k_L',k_L"}.
\end{eqnarray}

\subsection*{All equations}
Thus, we have all equations as follows:
\begin{eqnarray}
& &\left[ E_{k_{1}s}-E_2-\Sigma_1(E_{k_1})-\Sigma_3 (E_{k_1})
-Z_{k_1}|V_{k_1}|^2  \right] \nu^{(21)}_{k_{1}s}
-\Sigma_3 (E_{k_1})[\nu^{(41)}_{k_1s}]
=  0,  
\label{all_1}\\ 
& &\left[E_{k_{3}s}-E_2
-\Sigma_1(E_{k_3}) -\Sigma_3(E_{k_3})-Z_{k_3}|V_{k_3}|^2 \right]\nu^{(23)}_{k_3s}
-[\Sigma_3(E_{k_3})+Z_{k_3}|V_{k_3}|^2] [\nu^{(43)}_{k_3s}]= 0, 
\label{all_2}\\ 
& &\left[ E_{k_{5}s}-E_2 -\Sigma_1(E_{k_5}) -\Sigma_3(E_{k_5} ) \right]\nu^{(25)}_{k_5s}
-\Sigma_3(E_{k_5} )[\nu^{(45)}_{k_5s}] = 0,  
\label{all_3}\\
& &\left[E_{k_{1}s}-E_4-\Sigma_5(E_{k_1})-\Sigma_3(E_{k_1})\right]\nu^{(41)}_{k_{1}s}
-\Sigma_3(E_{k_1})[\nu^{(21)}_{k_1s}] = 0,  
\label{all_4}\\
& &\left[E_{k_{3}s}-E_4-\Sigma_5(E_{k_3})-\Sigma_3(E_{k_3})-Z_{k_3}|V_{k_3}|^2\right]\nu^{(43)}_{k_{3}s}
-[\Sigma_3(E_{k_3})+Z_{k_3}|V_{k_3}|^2][\nu^{(23)}_{k_3s}] = 0,
\label{all_5}\\
& &\left[E_{k_{5}s}-E_4-\Sigma_5(E_{k_5})-\Sigma_3(E_{k_5})-Z_{k_5}|V_{k_5}|^2 \right]\nu^{(45)}_{k_{5}s} 
-\Sigma_3(E_{k_5})[\nu^{(25)}_{k_5s}]  =0,
\label{all_6}\\
& &
|V_{k_5}|^2 
|\nu^{(45)}_{k_5s}|^2
( Z_{k_5}^2+\frac{L^2}{2u_k^2} )=1, \label{all_7}\\
 & & |V_{k_3}|^2 
|\nu^{(23)}_{k_3s}+\nu^{(43)}_{k_3s}|^2
( Z_{k_3}^2+\frac{L^2}{2u_k^2} )=1, \label{all_8}\\
& & |V_{k_1}|^2 
|\nu^{(21)}_{k_1s}|^2
( Z_{k_1}^2+\frac{L^2}{2u_k^2} )=1. \label{all_9}
\end{eqnarray}
In order to make it easier to see the structure of the equations, we use the following definitions
\begin{eqnarray}
e_1 &\equiv& E_{k_{1}s}-E_2-s_{11}-s_{31}, \\
e_2 &\equiv& E_{k_{3}s}-E_2-s_{13}-s_{33}, \\
e_3 &\equiv& E_{k_{5}s}-E_2-s_{15}-s_{35}, \\
e_4 &\equiv& E_{k_{1}s}-E_4-s_{51}-s_{31}, \\
e_5 &\equiv& E_{k_{3}s}-E_4-s_{53}-s_{33}, \\
e_6 &\equiv& E_{k_{5}s}-E_4-s_{55}-s_{35}. 
\end{eqnarray}
We also introduce the following notations:
\begin{eqnarray}
& &
x_1\equiv \nu^{(21)}_{k_1s}, \
x_2\equiv \nu^{(23)}_{k_3s}, \
x_3\equiv \nu^{(25)}_{k_5s},  \nnm\\
& &
y_1\equiv \nu^{(41)}_{k_1s}, \
y_2\equiv \nu^{(43)}_{k_3s}, \
y_3\equiv \nu^{(45)}_{k_5s}, \nnm \\
& &
z_1 \equiv Z_{k_1}|V_{k_1}|^2, \
z_3 \equiv Z_{k_3}|V_{k_3}|^2, \
z_5 \equiv Z_{k_5}|V_{k_5}|^2. \nnm 
\end{eqnarray}

Then, Eqs.(\ref{all_1})-(\ref{all_6}) are expressed as following:
\begin{eqnarray}
(e_1-z_1) x_1 -s_{31}y_1  &=& 0,  \\
(e_2-z_3) x_2 -(s_{33}+z_3)y_2 &=& 0,  \\
e_3 x_3  -s_{35} y_3  &=& 0, \\
e_4 y_1  -s_{31}x_1 &=& 0, \\
(e_5-z_3) y_2  -(s_{33}+z_3) x_2&=&0, \\
(e_6-z_5) y_3  -s_{35} x_3&=&0. 
\end{eqnarray}
By solving these equations, we have
\begin{eqnarray}
y_1&=&\frac{e_1-z_1}{s_{31}} x_1= \frac{s_{31}}{e_4} x_1,  \\
y_2&=&\frac{e_2-z_3}{(s_{33}+z_3)} x_2 =\frac{e_5+s_{33}}{e_2+s_{33}} x_2, \\
y_3&=&\frac{s_{35}}{e_6-z_5}=\frac{e_3}{s_{35}} x_3.   
\end{eqnarray}
From the second equations from the right, we have
\begin{eqnarray}
z_1 &=& e_1 -\frac{s_{31}^2}{e_4}, \\
z_5 &=& e_6 -\frac{s_{35}^2}{e_3}, \\
z_3 &=& \frac{e_2e_5-s_{33}^2}{e_2+e_5+2s_{33}}.
\end{eqnarray}
From Eqs.(\ref{all_6})-(\ref{all_9}), we have
\begin{eqnarray}  
|x_1|^2&=&\frac{1}{|V_{k_1}|^2 (Z_{k_1}^2+\frac{L^2}{2u_k^2})} 
=\frac{|V_{k_1}|^2}{ (Z_{k_1}^2|V_{k_1}|^4+\frac{L^2|V_{k_1}|^4}{2u_k^2})} 
=\frac{|V_{k_1}|^2}{ (e_1 -\frac{s_{31}^2}{e_4})^2+\Gamma_1^2}, \\
|y_3|^2&=&\frac{1}{|V_{k_5}|^2 ( Z_{k_5}^2+\frac{L^2}{2u_k^2} )} 
 =\frac{|V_{k_5}|^2}{ ( Z_{k_5}^2|V_{k_5}|^4+\frac{L^2|V_{k_5}|^4}{2u_k^2} )} 
 =\frac{|V_{k_5}|^2}{ (e_6 -\frac{s_{35}^2}{e_3})^2+\Gamma_5^2 }, \\
|x_2+y_2|^2
&=&\frac{1}{|V_{k_3}|^2 ( Z_{k_3}^2+\frac{L^2}{2u_k^2} )} 
=\frac{|V_{k_3}|^2 }{Z_{k_3}^2|V_{k_3}|^4 +\frac{L^2|V_{k_3}|^4}{2u_k^2} } 
=\frac{|V_{k_3}|^2 }{(\frac{e_2e_5-s_{33}^2}{e_2+e_5+2s_{33}})^2 +\Gamma_3^2}.
\label{x2+y2}
\end{eqnarray}
where
\begin{eqnarray}
\Gamma_1^2 &=& \frac{L^2|V_{k_1}|^4}{2u_k^2}, \
\Gamma_3^2 = \frac{L^2|V_{k_3}|^4}{2u_k^2}, \ 
\Gamma_5^2 = \frac{L^2|V_{k_5}|^4}{2u_k^2}.
\end{eqnarray}

We also have
\begin{eqnarray}
z_3+s_{33}&=&\frac{e_2e_5-s_{33}^2}{e_2+e_5+2s_{33}}+s_{33}
 =\frac{(e_2+s_{33})(e_5+s_{33})}{e_2+e_5+2s_{33}},
\nnm\\
e_5-z_3&=& e_5-\frac{e_2e_5-s_{33}^2}{e_2+e_5+2s_{33}}
=\frac{(e_5 +s_{33})^2}{e_2+e_5+2s_{33}}.
\nnm
\end{eqnarray}
Thus, we have
\begin{eqnarray}
x_2&=&\frac{e_5-z_3}{z_3+s_{33}}y_2 =\frac{(e_5 +s_{33})}{(e_2+s_{33})}y_2. \nnm 
\end{eqnarray}
From equation(\ref{x2+y2})
\begin{eqnarray}
|x_2|^2 |\frac{e_2+e_5+2s_{33}}{e_2+s_{33}}|^2
=\frac{|V_{k_3}|^2}{ \left( \frac{e_2e_5-s_{33}^2}{e_2+e_5+2s_{33}} \right)^2 +\Gamma_3^2}. \nnm
\end{eqnarray}
Thus
\begin{eqnarray}
|x_2|^2 
&=&\frac{|V_{k_3}|^2(e_2+s_{33})^2}{ \left( e_2e_5-s_{33}^2\right)^2 
+\Gamma_3^2 (e_2+e_5+2s_{33})^2},
\nnm \\
|y_2|^2 
&=&\frac{|V_{k_3}|^2(e_5+s_{33})^2}{ \left( e_2e_5-s_{33}^2\right)^2 
+\Gamma_3^2 (e_2+e_5+2s_{33})^2}.
\nnm 
\end{eqnarray}

For later use, we have derived the following equations:
\begin{eqnarray}
x_1x_1^*+y_1^*y_1&=&(1+(\frac{s_{31}}{e_4})^2)x_1^*x_1
=\frac{|V_{k_1}|^2(e_4^2+s_{31}^2)}{ (e_1e_4 -s_{31}^2)^2+e_4^2\Gamma_1^2},   
\nnm \\
x_3x_3^*+y_3^*y_3&=&((\frac{s_{35}}{e_3})^2+1)y_3^*y_3
=\frac{|V_{k_5}|^2(e_3^2+s_{35}^2)}{ (e_3e_6 -s_{35}^2)^2+e_3^2\Gamma_5^2 }, 
\nnm\\
|x_1x_2^*+y_1^*y_2|^2&=&(1+\frac{s_{31}}{e_4}\frac{e_5+s_{33}}{e_2+s_{33}})^2|x_1|^2|x_2|^2
\nnm \\
&=&
\frac{|V_{k_1}|^2|V_{k_3}|^2 (e_4(e_2+s_{33})+s_{31}(e_5+s_{33}))^2}{
[ (e_1e_4 -s_{31}^2)^2+e_4^2\Gamma_1^2][( e_2e_5-s_{33}^2)^2+\Gamma_3^2 (e_2+e_5+2s_{33})^2]},
\nnm \\
|x_1x_3^*+y_1^*y_3|^2&=&(\frac{s_{35}}{e_3}+\frac{s_{31}}{e_4})^2|x_1|^2|y_3|^2 \nnm \\
&=&
 \frac{|V_{k_1}|^2|V_{k_5}|^2[s_{35}e_4+ s_{31}e_3]^2}{ 
[(e_1e_4 -s_{31}^2)^2+e_4^2\Gamma_1^2]
[(e_6e_3 -s_{35}^2)^2+e_3^2\Gamma_5^2]}, 
\nnm \\
|x_2x_3^*+y_2^*y_3|^2&=&(\frac{s_{35}}{e_3}+\frac{e_5+s_{33}}{e_2+s_{33}})^2|x_2|^2|y_3|^2 \nnm \\
&=&
\frac{|V_{k_3}|^2|V_{k_5}|^2[s_{35}(e_2+s_{33})+e_3(e_5+s_{33})]^2 }{ 
\left[( e_2e_5-s_{33}^2)^2 +\Gamma_3^2 (e_2+e_5+2s_{33})^2\right]
[(e_6e_3 - s_{35}^2)^2+e_3^2\Gamma_5^2]}. \nnm
\end{eqnarray}


\section*{Kubo formula}
Here, we derive the conductance of the two QDs and three channels using the Kubo formula~\cite{Kubo}.
We calculate the total current consisting of the three currents:
\begin{equation}
J=\sum_{i=1,3,5}J_i.
\end{equation}
First we calculate the current-current correlation function in
\begin{eqnarray}
\Phi_{yy}^{\rm R} [t] &=&-\frac{i}{\hbar V}\theta(t) 
\la j_y(t)j_y(0) -j_y(0)j_y(t)\ra, 
\end{eqnarray}
where
\begin{eqnarray}
\la j_y(t)j_y(0) \ra &=&\la (\sum_{i=1,3,5}J_i(t))(\sum_{j=1,3,5}J_j(0))\ra \nnm\\
&=&\sum_{i,j=1,3,5}\sum_{k_i} \sum_{k_j'} (e\hbar/m)^2 k_ik_j'
\la c_{k_is}^\dagger (t) c_{k_is}(t) c_{k_j's}^\dagger c_{k_j's}\ra. 
\end{eqnarray}
The Bloch-De Dominicis's theorem is used here:
\begin{eqnarray}
& & \la c_{k_is}^\dagger (t) c_{k_is}(t) c_{k_j's}^\dagger c_{k_j's} \ra 
=
 \la c_{k_is}^\dagger (t) c_{k_is}(t)\ra \la c_{k_j's}^\dagger  c_{k_j's}\ra
+\la c_{k_is}^\dagger (t) c_{k_j's}\ra\la c_{k_is}(t)c_{k_j's}^\dagger \ra. \nnm
\end{eqnarray}
Then
\begin{eqnarray}
\la j_y(t)j_y(0) -j_y(0)j_y(t)  \ra 
&=&\sum_{i,j=1,3,5}\sum_{k_i} \sum_{k_j'} (e\hbar/m)^2 k_ik_j'
(\la c_{k_is}^\dagger (t) c_{k_j's}\ra\la c_{k_is}(t)c_{k_j's}^\dagger \ra 
-\la c_{k_is}^\dagger  c_{k_j's}(t)\ra\la c_{k_is}c_{k_j's}^\dagger (t)\ra ).
\end{eqnarray}
Because
\begin{equation}
i \frac{d \alpha_{k_i}}{dt}=[\alpha_{k_i}, H]
=[\alpha_i \sum_{k_i'} E_{k_i}\alpha_{k_i}^\dagger \alpha_{k_i}]
=E_{k_i} \alpha_{k_i},
\end{equation}
we have
$  
\alpha_{k_i} (t) =\exp (-iE_{k_i} t),
$  
and 
\begin{eqnarray}
\la\alpha_{k''_1s}(t)\alpha_{k''_1s}^\dagger\ra
&=& (1-f_{k_i''})  \exp (-iE_{k_i''} t),
\\
\la\alpha_{k''_is}^\dagger(t)\alpha_{k''_is}\ra
&=& f_{k_i''}  \exp (iE_{k_i''} t).
\end{eqnarray} 
Thus, we can calculate the following relations:
\begin{eqnarray}
\la c_{k_is}^\dagger (t) c_{k_is}(t)\ra 
&=&
 \sum_{k_1'} |\eta^{(i1)}_{k_i,k_1'}|^2 f_{k'_1s}
+\sum_{k_3'} |\eta^{(i3)}_{k_i,k_3'}|^2 f_{k'_3s}
+\sum_{k_5'} |\eta^{(i5)}_{k_i,k_5'}|^2 f_{k'_5s},
\nnm \\
\la c_{k_is}^\dagger (t) c_{k_j's}\ra
&=& 
 \sum_{k_1''} \eta^{(i1)*}_{k_i,k_1''}\eta^{(j1)}_{k_j',k_1''} 
  f_{k''_1s} e^{iE_{k_1''}t}
+\sum_{k_3''} \eta^{(i3)*}_{k_i,k_3''}\eta^{(j3)}_{k_j',k_3''}  
  f_{k''_3s} e^{iE_{k_3''}t}
+\sum_{k_5''} \eta^{(i5)*}_{k_i,k_5''} \eta^{(j5)}_{k_j',k_5''}
  f_{k''_5s} e^{iE_{k_5''}t},
\nnm \\
\la c_{k_is}(t)c_{k_j's}^\dagger \ra 
  &=&
 \sum_{k_1''} \eta^{(i1)}_{k_i,k_1''}\eta^{(j1)*}_{k_j',k_1''} 
  (1-f_{k''_1s}) e^{-iE_{k_1''}t}
+\sum_{k_3''} \eta^{(i3)}_{k_i,k_3''}\eta^{(j3)*}_{k_j',k_3''}  
  (1-f_{k''_3s}) e^{-iE_{k_3''}t}
+\sum_{k_5''} \eta^{(i5)}_{k_i,k_5''} \eta^{(j5)*}_{k_j',k_5''}
  (1-f_{k''_5s}) e^{-iE_{k_5''}t}. 
\nnm
\end{eqnarray}
Let us think about
\begin{eqnarray}
& &\sum_{i=1,3,5}\sum_{j=1,3,5}
\sum_{l_1=1,3,5}\sum_{l_2=1,3,5}
\sum_{k_i,k_j'}k_ik_j' \sum_{k_1''} \eta^{(il_1)*}_{k_i,k_{l_1}''}\eta^{(jl_1)}_{k_j',k_{l_1}''} 
  f_{k''_{l_1}s} e^{iE_{k_{l_1}''}t}
\sum_{k_{l_2}'''} \eta^{(i{l_2})}_{k_i,k_{l_2}'''}\eta^{(j{l_2})*}_{k_j',k_{l_2}'''} 
  (1-f_{k'''_{l_2}s}) e^{-iE_{k_{l_2}'''}t}
\nnm \\
&\sim &  
\sum_{l_1=1,3,5}\sum_{l_2=1,3,5}
\sum_{k_{l_1}''} \sum_{k_{l_2}'''}
\left[ 
\sum_{i=1,3,5}(\sum_{k_i}k_F^i
\eta^{(i{l_1})*}_{k_i,k_{l_1}''}
\eta^{(i{l_2})}_{k_i,k_{l_2}'''})
\sum_{j=1,3,5}(\sum_{k_j'}k_F^j  
\eta^{(j{l_1})}_{k_j',k_{l_1}''} 
\eta^{(j{l_2})*}_{k_j',k_{l_2}'''} )
\right]   f_{k''_{l_1}s} e^{iE_{k_{l_1}''}t} (1-f_{k'''_{l_2}s}) e^{-iE_{k_{l_2}'''}t}
\nnm \\
&= & \sum_{k_{1}''} \sum_{k_{1}'''}
\left[x_1''^*x_1'''+y_1''^*y_1'''-\delta_{k_1''}^{k_1'''}]
[x_1''x_1'''^*+y_1''y_1'''^*-\delta_{k_1''}^{k_1'''}\right]  
 f_{k_1''} e^{iE_{k_1''}t} (1-f_{k_1'''}) e^{-iE_{k_1'''}t}
\nnm \\
&+&  \sum_{k_{1}''} \sum_{k_{3}'''}
\left[x_1''^*x_2'''+y_1''^*y_2'''][x_1''x_2'''^*+y_1''y_2'''^*\right]  
f_{k_1''} e^{iE_{k_1''}t} (1-f_{k_3'''}) e^{-iE_{k_3'''}t}
\nnm \\
&+&  \sum_{k_{1}''} \sum_{k_{5}'''}
\left[x_1''^*x_3'''+y_1''^*y_3'''][x_1''x_3'''^*+y_1''y_3'''^*\right]  
f_{k_1''} e^{iE_{k_1''}t} (1-f_{k_5'''}) e^{-iE_{k_5'''}t}
\nnm \\
&+ & \sum_{k_{3}''} \sum_{k_{1}'''}
\left[x_2''^*x_1'''+y_2''^*y_1'''][x_2''x_1'''^*+y_2''y_1'''^*\right]  
 f_{k_3''} e^{iE_{k_3''}t} (1-f_{k_1'''}) e^{-iE_{k_1'''}t}
\nnm \\
&+&  \sum_{k_{3}''} \sum_{k_{3}'''}
\left[x_2''^*x_2'''+y_2''^*y_2'''-\delta_{k_3''}^{k_3'''}]
[x_2''x_2'''^*+y_2''y_2'''^*-\delta_{k_3''}^{k_3'''}\right]  
f_{k_3''} e^{iE_{k_3''}t} (1-f_{k_3'''}) e^{-iE_{k_3'''}t}
\nnm \\
&+&  \sum_{k_{3}''} \sum_{k_{5}'''}
\left[x_2''^*x_3'''+y_2''^*y_3'''][x_2''x_3'''^*+x_2''y_3'''^*\right]  
f_{k_3''} e^{iE_{k_3''}t} (1-f_{k_5'''}) e^{-iE_{k_5'''}t}
\nnm \\
&+ & \sum_{k_{5}''} \sum_{k_{1}'''}
\left[x_3''^*x_1'''+y_3''^*y_1'''][x_3''x_1'''^*+y_3''y_1'''^*\right]  
 f_{k_5''} e^{iE_{k_5''}t} (1-f_{k_1'''}) e^{-iE_{k_1'''}t}
\nnm \\
&+&  \sum_{k_{5}''} \sum_{k_{3}'''}
\left[x_3''^*x_2'''+y_3''^*y_2'''][x_3''x_2'''^*+y_3''y_2'''^*\right]  
f_{k_5''} e^{iE_{k_5''}t} (1-f_{k_3'''}) e^{-iE_{k_3'''}t}
\nnm \\
&+&  \sum_{k_{5}''} \sum_{k_{5}'''}
\left[x_3''^*x_3'''+y_3''^*y_3'''-\delta_{k_3''}^{k_3'''}]
[x_3''x_3'''^*+y_3''y_3'''^*-\delta_{k_3''}^{k_3'''}\right]  
f_{k_5''} e^{iE_{k_5''}t} (1-f_{k_5'''}) e^{-iE_{k_5'''}t}.
\end{eqnarray}
We pick up one element: 
\begin{eqnarray}
& &\sum_{i=3}\sum_{j=3}
\sum_{l_1=3}\sum_{l_2=3}
\sum_{k_i,k_j'}k_ik_j' \sum_{k_1''} \eta^{(il_1)*}_{k_i,k_{l_1}''}\eta^{(jl_1)}_{k_j',k_{l_1}''} 
  f_{k''_{l_1}s} e^{iE_{k_{l_1}''}t}
\sum_{k_{l_2}'''} \eta^{(i{l_2})}_{k_i,k_{l_2}'''}\eta^{(j{l_2})*}_{k_j',k_{l_2}'''} 
  (1-f_{k'''_{l_2}s}) e^{-iE_{k_{l_2}'''}t}
\nnm \\
&=&  
\sum_{k_{3}''} \sum_{k_{3}'''}
(x_2''^*x_2'''+y_2''^*y_2'''-\delta_{k_3''}^{k_3'''})
(x_2''x_2'''^*+y_2''y_2'''^*-\delta_{k_3''}^{k_3'''})
 f_{k''_{3}s} e^{iE_{k_{3}''}t} (1-f_{k'''_{3}s}) e^{-iE_{k_{3}'''}t}
\nnm\\
&=&  
\sum_{k_{3}''} \sum_{k_{3}'''}
|x_2''|^2|x_2'''|^2+y_2''^*x_2''y_2'''x_2'''^*
+x_2''^*x_2'''y_2''y_2'''^*+|y_2''|^2|y_2'''|^2) f_{k''_{3}s} e^{iE_{k_{3}''}t} (1-f_{k'''_{3}s}) e^{-iE_{k_{3}'''}t}
\nnm \\
& &
+\sum_{k_{3}''} \sum_{k_{3}'''}
(-|x_2''|^2\delta_{k_3''}^{k_3'''}
-|y_2''|^2\delta_{k_3''}^{k_3'''}
-(|x_2''|^2+|y_2''|^2-\delta_{k_3''}^{k_3'''})\delta_{k_3''}^{k_3'''})
 f_{k''_{3}s} e^{iE_{k_{3}''}t} (1-f_{k'''_{3}s}) e^{-iE_{k_{3}'''}t}.
\end{eqnarray}
The $\delta_{k_3''}^{k_3'''}$ parts are cancelled, 
and we have the $x$-part and $y$-part given by
\begin{eqnarray}
& &\sum_{i=3}\sum_{j=3}
\sum_{l_1=3}\sum_{l_2=3}
\sum_{k_i,k_j'}k_ik_j' \sum_{k_1''} \eta^{(il_1)*}_{k_i,k_{l_1}''}\eta^{(jl_1)}_{k_j',k_{l_1}''} 
  f_{k''_{l_1}s} e^{iE_{k_{l_1}''}t}
\sum_{k_{l_2}'''} \eta^{(i{l_2})}_{k_i,k_{l_2}'''}\eta^{(j{l_2})*}_{k_j',k_{l_2}'''} 
  (1-f_{k'''_{l_2}s}) e^{-iE_{k_{l_2}'''}t}
\nnm \\
&\sim &  
\sum_{k_{3}''} \sum_{k_{3}'''}
(x_2''^*x_2'''+y_2''^*y_2'''-\delta_{k_3''}^{k_3'''})
(x_2''x_2'''^*+y_2''y_2'''^*-\delta_{k_3''}^{k_3'''})
 f_{k''_{3}s} e^{iE_{k_{3}''}t} (1-f_{k'''_{3}s}) e^{-iE_{k_{3}'''}t}
\nnm\\
&=&  
\sum_{k_{3}''} \sum_{k_{3}'''}
|x_2''^*x_2'''+y_2''^*y_2'''|^2
 f_{k''_{3}s} e^{iE_{k_{3}''}t} (1-f_{k'''_{3}s}) e^{-iE_{k_{3}'''}t}.
\end{eqnarray}
From this equation, we have 
\begin{eqnarray}
\la j_y(t)j_y(0) -j_y(0)j_y(t)  \ra 
&=&\sum_{i,j=1,3,5}\sum_{k_i} \sum_{k_j'} (e\hbar/m)^2 k_ik_j'
(\la c_{k_is}^\dagger (t) c_{k_j's}\ra\la c_{k_is}(t)c_{k_j's}^\dagger \ra 
-\la c_{k_is}^\dagger  c_{k_j's}(t)\ra\la c_{k_is}c_{k_j's}^\dagger (t)\ra )
\nnm \\
&=&  
\frac{(e\hbar k_F)^2}{m^2}\sum_{k_{3}''} \sum_{k_{3}'''}
|x_2''^*x_2'''+y_2''^*y_2'''|^2
 (f_{k''_{3}s}-f_{k'''_{3}s}) e^{i[E_{k_{3}''}-E_{k_{3}'''}]/\hbar t}.
\end{eqnarray}
Then
\begin{eqnarray}
\Phi_{yy}^{\rm R} (\omega) &=&
\int_{-\infty}^{\infty} \frac{-i}{\hbar V} \theta(t) 
\la j_y (t) j_y (0)-j_y(0)j_y(t) \ra e^{i\omega t} dt
\nnm \\
&=& \frac{-i}{\hbar V} 
\sum_{k_{3}''} \sum_{k_{3}'''}
\left[
\frac{i}{\omega-(E_{k_{3}'''}-E_{k_{3}''})/\hbar}+\pi \delta(\omega-(E_{k_{3}'''}-E_{k_{3}''})/\hbar)
\right]
\frac{(e\hbar k_F)^2}{m^2}
|x_2''^*x_2'''+y_2''^*y_2'''|^2
 (f_{k''_{3}s}-f_{k'''_{3}s}). 
\nnm 
\end{eqnarray}
The 1-st term disappear because of $ (f_{k''_{3}s}-f_{k'''_{3}s})$.
In addition, we use the density of states (DOS) of a unit volume as follows:
\begin{eqnarray}
\sum_i \rightarrow \sum_k^{v} &\rightarrow & \frac{V}{L} \int \frac{d{\vec k}}{(2\pi)^3}  
\rightarrow \frac{V}{L} \int dE D(E).  
\end{eqnarray}
So we have to calculate
\begin{eqnarray}
\lefteqn{ 
\Phi_{yy}^{\rm R'} (\omega) 
= \frac{1}{\hbar V} 
\sum_{k_{3}''} \sum_{k_{3}'''}
\left[\pi \delta(\omega-(E_{k_{3}'''}-E_{k_{3}''})/\hbar)\right]
\frac{(e\hbar k_F)^2}{m^2}
|x_2''^*x_2'''+y_2''^*y_2'''|^2
 (f_{k''_{3}s}-f_{k'''_{3}s}) }
\nnm\\
&=& \frac{V}{L^2}\frac{\pi(e\hbar k_F)^2}{m^2}
\int_{-\infty}^{\infty}D(E_{k_{3}''}) D(E_{k_{3}''}+\hbar\omega)dE_{k_{3}''} 
|x_2^*(E_{k_{3}''})x_2(E_{k_{3}''}+\hbar\omega)+y_2^*(E_{k_{3}''})y_2(E_{k_{3}''}+\hbar\omega)|^2
 (f(E_{k_{3}''})-f((E_{k_{3}''}+\hbar\omega))). \nnm
\end{eqnarray}
Let us use the following equation for an arbitrary function $A(E,E+\omega)$;
\begin{eqnarray}
\lefteqn{
\frac{\partial }{\partial \omega}\int A(E,E+\omega)(f(E)-f(E+\omega)) dE  
}\nnm \\
&=&
 \int \frac{\partial A(E, E+\omega)}{\partial \omega} (f(E)-f(E+\omega))dE  
+\int A(E,E+\omega)\frac{\partial f(E)-f(E+\omega) }{\partial \omega}dE. 
\end{eqnarray}
In this equation, the 1st term can be neglected. Then we have
\begin{eqnarray}
\lefteqn{
g_{yy}
\Rightarrow 
\frac{\partial \Phi_{yy}^{\rm R'} (\omega)}{\partial \omega}|_{\omega \rightarrow 0}  
}
\nnm \\
&=& \frac{V\pi(e\hbar k_F)^2}{L^2m^2} \! \!
\int_{-\infty}^{\infty}\! dE_{k_{3}''} D(E_{k_{3}''}) D(E_{k_{3}''}+\hbar\omega)  
|x_2^*(E_{k_{3}''})x_2(E_{k_{3}''}+\hbar\omega)+y_2^*(E_{k_{3}''})y_2(E_{k_{3}''}+\hbar\omega)|^2
\frac{\partial  (f(E_{k_{3}''})-f((E_{k_{3}''}+\hbar\omega))) }{\partial \omega}
\nnm \\
&=& \frac{V\pi(e\hbar k_F)^2}{L^2m^2}D(E_{k_F})^2
|x_2^*(E_{k_F})x_2(E_{k_F})+y_2^*(E_{k_F})y_2(E_{k_F})|^2\hbar
\nnm \\
&=& \frac{V\hbar\pi(e\hbar k_F)^2}{L^2m^2}D(E_{k_F})^2
\frac{|V_{k_F}|^4
\left[(e_2+s_{33})^2+ (e_5+s_{33})^2)\right]^2 }
{\left[ ( e_2e_5-s_{33}^2)^2 +\Gamma_3^2 (e_2+e_5+2s_{33})^2 \right]^2 } 
\nnm \\
&=& \frac{V\hbar\pi(e\hbar k_F)^2}{L^2m^2}D(E_{k_F})^2
\frac{|V_{k_F}|^4
\left[(E_{k_F}-E_2)^2+ (E_{k_F}-E_4)^2)\right]^2 }
{\left[[(E_{k_F}-E_2)(E_{k_F}-E_4)+(2E_{k_F}-E_2-E_4)-s_{33}]^2 
+\Gamma_3^2 (2E_{k_F}-E_2-E_4)^2 \right]^2 },
\end{eqnarray}
where we use $x=\hbar \omega$
\begin{equation}
\frac{\partial f(\hbar \omega)}{\partial \omega}
=\frac{\partial f(x)}{\partial x}\frac{\partial (\hbar \omega)}{\partial \omega}
=-\delta(x-\mu_F) \hbar
=-\delta(\hbar \omega-\mu_F)\hbar.
\end{equation}
Thus, we have the general conductance formula:
\begin{eqnarray}
g_{yy}
&\Rightarrow &
\frac{\partial \Phi_{yy}^{\rm R'} (\omega)}{\partial \omega}|_{\omega \rightarrow 0}  
\nnm \\
&= & \frac{V\hbar\pi(e\hbar k_F)^2}{L^2m^2} D(E_{k_F})^2 \{
    |x_1^*(E_{k_F})x_1(E_{k_F})+y_1^*(E_{k_F})y_1(E_{k_F})|^2 \nnm \\
&+& |x_2^*(E_{k_F})x_2(E_{k_F})+y_2^*(E_{k_F})y_2(E_{k_F})|^2 \nnm \\
&+& |x_3^*(E_{k_F})x_3(E_{k_F})+y_3^*(E_{k_F})y_3(E_{k_F})|^2 \nnm \\
&+& 2|x_1^*(E_{k_F})x_2(E_{k_F})+y_1^*(E_{k_F})y_2(E_{k_F})|^2 \nnm \\
&+& 2|x_1^*(E_{k_F})x_3(E_{k_F})+y_1^*(E_{k_F})y_3(E_{k_F})|^2 \nnm \\
&+& 2|x_2^*(E_{k_F})x_3(E_{k_F})+y_2^*(E_{k_F})y_3(E_{k_F})|^2 
\}
\nnm \\
&=& \frac{V\hbar\pi(e\hbar k_F)^2}{L^2m^2} D(E_{k_F})^2 \Biggl\{
    \frac{|V_{k_1}|^4(e_4^2+s_{31}^2)^2}{[(e_1e_4 -s_{31}^2)^2+e_4^2\Gamma_1^2]^2}
  + \frac{|V_{k_5}|^4(e_3^2+s_{35}^2)^2}{[(e_3e_6 -s_{35}^2)^2+e_3^2\Gamma_5^2]^2} \nnm \\
&+& \frac{|V_{k_F}|^4 [(e_2+s_{33})^2+ (e_5+s_{33})^2)]^2 }
{\left[ ( e_2e_5-s_{33}^2)^2 +\Gamma_3^2 (e_2+e_5+2s_{33})^2 \right]^2 }  
\nnm \\
&+& \frac{2|V_{k_1}|^2|V_{k_3}|^2 (e_4(e_2+s_{33})+s_{31}(e_5+s_{33}))^2}{
[ (e_1e_4 -s_{31}^2)^2+e_4^2\Gamma_1^2][( e_2e_5-s_{33}^2)^2+\Gamma_3^2 (e_2+e_5+2s_{33})^2]}\nnm \\
&+& \frac{2|V_{k_1}|^2|V_{k_5}|^2[s_{35}e_4+ s_{31}e_3]^2}{ 
[(e_1e_4 -s_{31}^2)^2+e_4^2\Gamma_1^2]
[(e_6e_3 -s_{35}^2)^2+e_3^2\Gamma_5^2]} 
 \nnm \\
&+& \frac{2|V_{k_3}|^2|V_{k_5}|^2[s_{35}(e_2+s_{33})+e_3(e_5+s_{33})]^2 }{ 
\left[( e_2e_5-s_{33}^2)^2 +\Gamma_3^2 (e_2+e_5+2s_{33})^2\right]
[(e_6e_3 - s_{35}^2)^2+e_3^2\Gamma_5^2]} 
\Biggr\}.
\end{eqnarray}
Here, we have used the following equations:
\begin{eqnarray}
x_1x_1^*+y_1^*y_1
&=&\frac{|V_{k_1}|^2(e_4^2+s_{31}^2)}{ (e_1e_4 -s_{31}^2)^2+e_4^2\Gamma_1^2}   
\Rightarrow
\frac{|V_{k_1}|^2(e_5^2+s_{31}^2)}{ (e_2e_5 -s_{31}^2)^2+e_5^2\Gamma_1^2},   
\nnm \\
x_3x_3^*+y_3^*y_3&=&
=\frac{|V_{k_5}|^2(e_3^2+s_{35}^2)}{ (e_3e_6 -s_{35}^2)^2+e_3^2\Gamma_5^2 } 
\Rightarrow\frac{|V_{k_5}|^2(e_2^2+s_{35}^2)}{ (e_2e_5 -s_{35}^2)^2+e_2^2\Gamma_5^2 }, 
\nnm\\
|x_1x_2^*+y_1^*y_2|^2
&=&
\frac{|V_{k_1}|^2|V_{k_3}|^2 (e_4(e_2+s_{33})+s_{31}(e_5+s_{33}))^2}{
[ (e_1e_4 -s_{31}^2)^2+e_4^2\Gamma_1^2][( e_2e_5-s_{33}^2)^2+\Gamma_3^2 (e_2+e_5+2s_{33})^2]}
\nnm \\
&\Rightarrow&
\frac{|V_{k_1}|^2|V_{k_3}|^2 (e_5(e_2+s_{33})+s_{31}(e_5+s_{33}))^2}{
[ (e_2e_5 -s_{31}^2)^2+e_5^2\Gamma_1^2][( e_2e_5-s_{33}^2)^2+\Gamma_3^2 (e_2+e_5+2s_{33})^2]},
\nnm \\
|x_1x_3^*+y_1^*y_3|^2
&=&
\frac{|V_{k_1}|^2|V_{k_5}|^2[s_{35}e_4+ s_{31}e_3]^2}{ 
[(e_1e_4 -s_{31}^2)^2+e_4^2\Gamma_1^2]
[(e_6e_3 -s_{35}^2)^2+e_3^2\Gamma_5^2]} 
\nnm \\
&\Rightarrow&
\frac{|V_{k_1}|^2|V_{k_5}|^2[s_{3F}(e_5+ e_2)]^2}{ 
[(e_2e_5 -s_{31}^2)^2+e_5^2\Gamma_1^2]
[(e_2e_5 -s_{35}^2)^2+e_2^2\Gamma_5^2]},  
\nnm \\
|x_2x_3^*+y_2^*y_3|^2
&=&
\frac{|V_{k_3}|^2|V_{k_5}|^2[s_{35}(e_2+s_{33})+e_3(e_5+s_{33})]^2 }{ 
\left[( e_2e_5-s_{33}^2)^2 +\Gamma_3^2 (e_2+e_5+2s_{33})^2\right]
[(e_6e_3 - s_{35}^2)^2+e_3^2\Gamma_5^2]}
\nnm \\
&\Rightarrow&
\frac{|V_{k_3}|^2|V_{k_5}|^2[s_{35}(e_2+s_{33})+e_2(e_5+s_{33})]^2 }{ 
\left[( e_2e_5-s_{33}^2)^2 +\Gamma_3^2 (e_2+e_5+2s_{33})^2\right]
[(e_2e_5 - s_{35}^2)^2+e_2^2\Gamma_5^2]}.
\nnm 
\end{eqnarray}

\subsection*{Detail of calculations}
In the above derivations, we have used the following equations:
\begin{eqnarray}
\sum_{k_1}
\eta^{(11)}_{k_1,k_1''}\eta^{(13)*}_{k_1,k_3'''}
+\sum_{k_3}
\eta^{(31)}_{k_3,k_1''}\eta^{(33)*}_{k_3,k_3'''}
+\sum_{k_5}
\eta^{(51)}_{k_5,k_1''}\eta^{(53)*}_{k_5,k_3'''}
&=&-x_1''x_2'''^*-y_1''y_2'''^*,
\nnm\\
 \sum_{k_1}
 \eta^{(11)}_{k_1,k_1''}\eta^{(15)*}_{k_1,k_5'''}
+\sum_{k_3}
 \eta^{(31)}_{k_5,k_1''}\eta^{(35)*}_{k_3,k_5'''}
+\sum_{k_5}
\eta^{(51)}_{k_5,k_1''}\eta^{(55)*}_{k_5,k_5'''}
&=& - x_1'' x_3'''^* -y_1''y_3'''^*,
\nnm\\
\sum_{k_1}
\eta^{(13)}_{k_1,k_3''}\eta^{(11)*}_{k_1,k_1'''}
+\sum_{k_3}
\eta^{(33)}_{k_3,k_3''}\eta^{(31)*}_{k_3,k_1'''}
+\sum_{k_5}
\eta^{(53)}_{k_5,k_3''}\eta^{(51)*}_{k_5,k_1'''}
&=&
-x_2''x_1'''^*-y_2''y_1'''^*,
\nnm \\
\sum_{k_1}
\eta^{(11)}_{k_1,k_1''}\eta^{(11)*}_{k_1,k_1'''}
+\sum_{k_3}
\eta^{(31)}_{k_3,k_1''}\eta^{(31)*}_{k_3,k_1'''}
+\sum_{k_5}
\eta^{(51)}_{k_5,k_1''}\eta^{(51)*}_{k_5,k_1'''}
&=&
-x_1''x_1'''^* -y_1''y_1'''^*+\delta_{k_1'''}^{k_1''},
\nnm\\
\sum_{k_1}
\eta^{(13)}_{k_1,k_3''}\eta^{(13)*}_{k_1,k_3'''}
+\sum_{k_5}
\eta^{(53)}_{k_5,k_3''}\eta^{(53)*}_{k_5,k_3'''}
+\sum_{k_3}
\eta^{(33)}_{k_3,k_3''}\eta^{(33)*}_{k_3,k_3'''}
&=&-x_2''x_2'''^*-y_2''y_2'''^*+\delta_{k_3'''}^{k_3''},
\nnm \\
\sum_{k_1}
\eta^{(13)}_{k_1,k_3''} \eta^{(15)*}_{k_1,k_5'''}
+\sum_{k_3}
\eta^{(33)}_{k_3,k_3''}\eta^{(35)*}_{k_3,k_5'''}
+\sum_{k_5}
\eta^{(53)}_{k_5,k_3''}\eta^{(55)*}_{k_5,k_5'''}
&=&-x_2''x_3'''^*-y_2''y_3'''^*,
\nnm \\
\sum_{k_1}
\eta^{(15)}_{k_1,k_5''}
\eta^{(11)*}_{k_1,k_1'''}
+\sum_{k_3}
\eta^{(35)}_{k_1,k_5''}\eta^{(31)*}_{k_3,k_1'''}
+\sum_{k_5}
\eta^{(55)}_{k_5,k_5''}\eta^{(51)*}_{k_5,k_1'''}
&=&-x_3''x_1'''^*-x_3'' x_1'''^* ,
\nnm \\
 \sum_{k_1}
\eta^{(15)}_{k_1,k_5''}\eta^{(13)*}_{k_1,k_3'''}
+\sum_{k_3}
\eta^{(35)}_{k_3,k_5''}\eta^{(33)*}_{k_3,k_3'''}
+\sum_{k_5}
\eta^{(55)}_{k_5,k_5''} \eta^{(53)*}_{k_5,k_3'''}
&=&-x_3''x_2'''^*-y_3''y_2'''^* ,
\nnm \\
\sum_{k_1}
\eta^{(15)}_{k_1,k_5''}\eta^{(15)*}_{k_1,k_5'''}
+\sum_{k_3}
\eta^{(35)}_{k_3,k_5''}\eta^{(35)*}_{k_3,k_5'''}
+\sum_{k_1}
\eta^{(55)}_{k_5,k_5''} \eta^{(55)*}_{k_5,k_5'''}
&=&-x_3''x_3'''-y_3''y_3'''+\delta_{k_5'''}^{k_5''}. \nnm
\end{eqnarray}
These equations are derived by using Eqns.(\ref{eta1})$\sim$(\ref{eta9}).
For example, the case of \{(11)(13)+(31)(33)+(51)(53)\} is given as follows:
\begin{eqnarray}
\sum_{k_1}
\eta^{(11)}_{k_1,k_1''} \eta^{(13)*}_{k_1,k_3'''}
&=&\sum_{k_1}(-\frac{V_{k_1}}{E_{k_1}-E_{k_1''}}\nu^{(21)}_{k_1''s} 
+\delta_{k_1}^{k_1''}Z_{k_1}V_{k_1} \nu^{(21)}_{k_1s})
(-\frac{V_{k_1}^*}{E_{k_1}-E_{k_3'''}}\nu^{(23)*}_{k_3'''s})
\nnm \\
&=&
\frac{1}{E_{k_1''}-E_{k_3'''}}
\left(
\Sigma_1 (E_{k_3'''})-\Sigma_1 (E_{k_1''}) 
-Z_{k_1''}|V_{k_1''}|^2
\right)\nu^{(21)}_{k''_1s}\nu^{(23)*}_{k_3'''s}
\nnm \\
&=&
\frac{1}{E_{k_1''}-E_{k_3'''}}
\left( s_{13}'''-s_{11}''-z_1'' \right) x_1''x_2'''^*,
\nnm \\
\sum_{k_1}
\eta^{(13)}_{k_1,k_3''}
\eta^{(11)*}_{k_1,k_1'''}
&=&\sum_{k_1}
(-\frac{V_{k_1}}{E_{k_1}-E_{k_3''}}\nu^{(23)}_{k_3''s})
(-\frac{V_{k_1}^*}{E_{k_1}-E_{k_1'''}}\nu^{(21)*}_{k_1'''s} 
+\delta_{k_1}^{k_1'''}Z_{k_1}V_{k_1} \nu^{(21)*}_{k_1s})
\nnm \\
&=&
\frac{1}{E_{k_1'''}-E_{k_3''}}
\left(
\Sigma_1 (E_{k_3''})-\Sigma_1 (E_{k_1'''}) 
-Z_{k_1'''}|V_{k_1'''}|^2
\right)\nu^{(23)}_{k_3''s}\nu^{(21)*}_{k'''_1s}
\nnm \\
&=&
\frac{1}{E_{k_1'''}-E_{k_3''}}
\left( s_{13}''-s_{11}'''-z_1''' \right)x_2''x_1*''',
\nnm \\
\sum_{k_3}
\eta^{(31)}_{k_3,k_1''}\eta^{(33)*}_{k_3,k_3'''}
&=&
\frac{1}{E_{k_3'''}-E_{k_1''}}
\left(
\Sigma_3 (E_{k_1''})-\Sigma_3 (E_{k_3'''}) 
-Z_{k_3'''}|V_{k_3'''}|^2
\right)[\nu^{(21)}_{k_1''s}+\nu^{(41)}_{k_1''s}][\nu^{(23)*}_{k'''_3s}+\nu^{(43)*}_{k'''_3s}]
\nnm \\
&=&
\frac{1}{E_{k_3'''}-E_{k_1''}}
\left( s_{31}''-s_{33}'''-z_{3}'''\right)[x_1''+y_1''][x_2'''^*+y_2'''^*],
\nnm \\
\sum_{k_5}
\eta^{(51)}_{k_5,k_1''}\eta^{(53)*}_{k_5,k_3'''}
&=&
\sum_{k_5}
\frac{V_{k_5}}{E_{k_5}-E_{k_1''}} \nu^{(41)}_{k_1''s}
\frac{V_{k_5}^*}{E_{k_5}-E_{k_3'''}}\nu^{(43)*}_{k_3'''s}
\nnm \\
&=&
\sum_{k_5}
\frac{|V_{k_5}|^2}{E_{k_1''}-E_{k_3'''}} 
(\frac{1}{E_{k_5}-E_{k_1''}}-\frac{1}{E_{k_5}-E_{k_3'''}})
\nu^{(41)}_{k_1''s}\nu^{(43)*}_{k_3'''s}
\nnm \\
&=&
\frac{1}{E_{k_1''}-E_{k_3'''}} 
(\Sigma_5(E_{k_3'''})-\Sigma_5(E_{k_1''}))
\nu^{(41)}_{k_1''s}\nu^{(43)*}_{k_3'''s}
\nnm \\
&=&
\frac{1}{E_{k_1''}-E_{k_3'''}} 
(s_{53}'''-s_{51}'')y_1''y_2'''*. \nnm
\end{eqnarray}
Thus,
\begin{eqnarray}
\lefteqn{
\left[
\sum_{k_1}
\eta^{(11)}_{k_1,k_1''}\eta^{(13)*}_{k_1,k_3'''}
+\sum_{k_3}
\eta^{(31)}_{k_3,k_1''}\eta^{(33)*}_{k_3,k_3'''}
+\sum_{k_5}
\eta^{(51)}_{k_5,k_1''}\eta^{(53)*}_{k_5,k_3'''}
\right]
(E_{k_1''}-E_{k_3'''}) }\nnm\\
&=&( s_{13}'''-s_{11}''-z_1'' ) x_1''x_2'''^*
-( s_{31}''-s_{33}'''-z_{3}''')[x_1''+y_1''][x_2'''^*+y_2'''^*]
+(s_{53}'''-s_{51}'')y_1''y_2'''^*
\nnm \\
&=&(-E_2-E_{k_1}'' +E_{k_3}'''-E_2'''  ) x_1''x_2'''^*
(E_{k_3}'''-E_4'''-E_{k_1}'')y_1''y_2'''^*
\nnm \\
&=&(-E_{k_1}'' +E_{k_3}''' )( x_1''x_2'''^*+y_1''y_2'''^*).
\end{eqnarray}
Then, we get 
\begin{eqnarray}
\sum_{k_1}
\eta^{(11)}_{k_1,k_1''}\eta^{(13)*}_{k_1,k_3'''}
+\sum_{k_3}
\eta^{(31)}_{k_3,k_1''}\eta^{(33)*}_{k_3,k_3'''}
+\sum_{k_5}
\eta^{(51)}_{k_5,k_1''}\eta^{(53)*}_{k_5,k_3'''}
=-x_1''x_2'''^*-y_1''y_2'''^*.
\nnm
\end{eqnarray}

\section*{The $s$-$d$ interaction under magnetic field}
The RKKY interaction is derived from the $s$-$d$ interaction.
Here, we would like to investigate the form of the $s$-$d$ interaction under a magnetic field following Ref.\cite{Coqblin}.
The unperturbed Hamiltonian is the Anderson Hamiltonian and the 
perturbation term is given by
\begin{eqnarray}
H_1 &=& \sum_{k_q m_q} (V_{k_q}c_{k_qm_q}^\dagger c_{m_q} +V_{k_q}^*c_{m_q}^\dagger c_{k_qm_q}).
\end{eqnarray}
Then the second order term is given by using the canonical transformation as follows:
\begin{eqnarray}
H_2&=&\frac{1}{2}\sum_{abc} \la b| H_1 |c\ra\la c| H_1|a\ra
\left( \frac{1}{e_a-e_c}+\frac{1}{e_b-e_c}\right),  
\end{eqnarray}
where $a$ and $b$ are the initial and final states, respectively, 
and $c$ is the intermediate state:
\begin{eqnarray}
e_a &\equiv& e_{km_{a1}} +e_{m_{a2}}, \\
e_b &\equiv& e_{k'm_{b1}}+e_{m_{b2}}, \\
e_c &\equiv& e_{d\uparrow}+e_{d\downarrow}+U,
\end{eqnarray}
the corresponding states are given by
\begin{eqnarray}
|a\ra&=& c_{k_{a1}m_{a1}}^\dagger c_{m_{a2}}^\dagger |0\ra, \
|b\ra= c_{k_{b1}m_{b1}}^\dagger c_{m_{b2}}^\dagger |0\ra, \\
|c_0\ra&=& c_{k_{c1}m_{c1}}^\dagger c_{k_{c2}m_{c2}}^\dagger |0\ra, \
|c_u\ra= c_{m_{c1}}^\dagger c_{m_{c2}}^\dagger |0\ra. 
\end{eqnarray}
When the intermediate state is empty, we have
\begin{eqnarray}
\la b | H_1 |c_0 \ra 
&=& 
\la 0 | c_{m_{b2}} c_{k_{b1}m_{b1}} 
\sum_{k_q m_q} (V_{k_q}c_{k_qm_q}^\dagger c_{m_q} +V_{k_q}^*c_{m_q}^\dagger c_{k_qm_q})
c_{k_{c1}m_{c1}}^\dagger c_{k_{c2}m_{c2}}^\dagger |0\ra \nnm \\
&=& 
-(V_{k_{c1}}^*\delta_{m_{b2}}^{m_{c1}}\delta_{m_{b1}}^{m_{c2}}
 \delta_{k_{b1}}^{k_{c2}}
 -V_{k_{c2}}^*\delta_{m_{b2}}^{m_{c2}}\delta_{m_{b1}}^{m_{c1}}
 \delta_{k_{b1}}^{k_{c1}}
 ),
 \\
\la c_0 | H_1 | a \ra 
&=& 
-(V_{k_{c1}}\delta_{m_{a2}}^{m_{c1}}\delta_{m_{a1}}^{m_{c2}}
 \delta_{k_{a1}}^{k_{c2}}
 -V_{k_{c2}}\delta_{m_{a2}}^{m_{c2}}\delta_{m_{a1}}^{m_{c1}}
 \delta_{k_{a1}}^{k_{c1}}
 ).
\end{eqnarray}
\begin{eqnarray}
H_2&=&
\sum_{abc} 
2|V_{k_{c1}}|^2
c_{k_{a1}-m_{a2}}^\dagger c_{k_{a1},-m_{a2}} n_{m_{a2}}   
 \left(
\frac{1}{e_{m_{a2}}-e_{k_{c1}m_{a2}}}\right) 
\nnm \\
&-&\sum_{abc} 
V_{k_{a1}}^*V_{k_{b1}}
c_{k_{b1}m_{a2}}^\dagger c_{k_{a1},-m_{a2}}
c_{-m_{a2}}^\dagger c_{m_{a2}}   
 \left(
 \frac{1}{e_{m_{a2}}-e_{k_{b1}m_{a2}}}
+\frac{1}{e_{-m_{a2}}-e_{k_{a1}-m_{a2}}}\right)  \nnm \\
&=&
\sum_{k} |V_{k}|^2 
\frac{1}{2} [
(1-4S_{kz}S_{z}) [F_{k\uparrow}+F_{k\downarrow}]
+2(S_z-S_{kz}) [F_{k\uparrow}-F_{k\downarrow}]
]\nnm \\
&-&\sum_{k} |V_{k}|^2 
2(S_k^xS^x+S_k^yS^y) [F_{k\uparrow}+F_{k\downarrow}] \nnm 
\nnm \\
&=&
-2 \sum_{k} |V_{k}|^2 [F_{k\uparrow}+F_{k\downarrow}] {\vec S}_k {\vec S} 
+ \sum_{k} |V_{k}|^2 [F_{k\uparrow}-F_{k\downarrow}](S_z-S_{kz}) 
+\frac{1}{2} \sum_{k} |V_{k}|^2  [F_{k\uparrow}+F_{k\downarrow}],
\end{eqnarray}
where we define:
\begin{eqnarray}
S_z&=&\frac{1}{2}(n_\uparrow -n_\downarrow), 
\ S_{kz}=\frac{1}{2}(n_{k\uparrow} -n_{k\downarrow}), \\
S_+&=&c_{\uparrow}^\dagger c_{\downarrow}, \  S_-=c_{\downarrow}^\dagger c_{\uparrow}, \\ 
S_{k+}&=&c_{k\uparrow}^\dagger c_{k\downarrow}, \  S_{k-}=c_{k\downarrow}^\dagger c_{k\uparrow}, \\ 
F_{k\pm}&=& \frac{1}{e_{\pm}-e_{k}}.
\end{eqnarray}


Thus, the magnetic fields enter all elements of x,y, and z elements of the couplings.
When the intermediate state is doubly occupied, we have
\begin{eqnarray}
\la b | H_1 |c_u \ra 
&=& 
\la 0 | c_{m_{b2}} c_{k_{b1}m_{b1}} 
\sum_{k_q m_q} (V_{k_q}c_{k_qm_q}^\dagger c_{m_q} +V_{k_q}^*c_{m_q}^\dagger c_{k_qm_q})
c_{m_{c1}}^\dagger c_{m_{c2}}^\dagger |0\ra \nnm \\
&=&  
V_{k_{b1}} 
(\delta_{m_{b2}}^{m_{c2}}\delta_{m_{b1}}^{m_{c1}}
-\delta_{m_{b2}}^{m_{c1}}\delta_{m_{b1}}^{m_{c2}}), \\
\la c_u | H_1 | a \ra 
&=& 
V_{k_{a1}}^*
(\delta_{m_{a2}}^{m_{c2}}\delta_{m_{a1}}^{m_{c1}}
-\delta_{m_{a2}}^{m_{c1}}\delta_{m_{a1}}^{m_{c2}}). 
\end{eqnarray}
Then, we have
\begin{eqnarray}
\lefteqn{ H_2
=\frac{1}{2}\sum_{abc} 
c_{k_{b1}m_{b1}}^\dagger c_{m_{b2}}^\dagger |0\ra
\la b| H_1 |c\ra\la c| H_1|a\ra \la 0|
c_{m_{a2}} c_{k_{a1}m_{a1}}  
\left( \frac{1}{e_a-e_c}+\frac{1}{e_b-e_c}\right)} \nnm\\
&=&\sum_{abc} 
|V_{k_{c1}}|^2 
c_{k_{c1}-m_{a2}}^\dagger  c_{k_{c1}-m_{a2}} 
c_{m_{a2}}^\dagger c_{m_{a2}}   
\left(
 \frac{1}{e_{k_{c1}-m_{a2}}-e_{-m_{a2}}}
+\frac{1}{e_{k_{c1}-m_{a2}}-e_{-m_{a2}}}\right) \nnm\\
&-&\sum_{abc} 
V_{k_{c1}} V_{k_{c2}}^*
c_{k_{c1}m_{a2}}^\dagger c_{k_{c2}-m_{a2}} 
c_{-m_{a2}}^\dagger 
c_{m_{a2}}   
\left(
 \frac{1}{e_{k_{c2}-m_{a2}}-e_{-m_{a2}}}
+\frac{1}{e_{k_{c1}m_{a2}}-e_{m_{a2}}} \right). 
\end{eqnarray}
In this case, we have similar form of the interaction between $\vec{S}$ and $\vec{S}_k$.



\end{document}